\documentclass[aps,prd,reprint,superscriptaddress,floatfix,nofootinbib,showkeys,showpacs,preprintnumbers]{revtex4-2}
\pdfoutput=1

\usepackage{tabularx}
\usepackage{xspace}
\usepackage{listings}
\usepackage{lineno}
\usepackage{bm}
\usepackage{amsmath}
\usepackage{amssymb}
\usepackage{rotating}
\usepackage[dvipsnames]{xcolor} % Heidi: svgnames seems to cause conflict
\usepackage{graphicx}
\usepackage{comment}
\usepackage{natbib}
\usepackage{hyperref}
\usepackage{amsmath}
\usepackage{enumitem}
\setlist{wide, labelwidth=!, labelindent=0pt}
\usepackage{multirow}
\hypersetup{    
  colorlinks      = true,
  linkcolor       = {black},
  linkbordercolor = {white},
  citecolor       = {black},
  citebordercolor = {white},
  urlcolor        = {blue},
  urlbordercolor  = {white},
}

\usepackage{soul} 
\usepackage{amsmath}
\usepackage{amssymb}
\usepackage{xspace}
\usepackage{xifthen}
\usepackage{float}
\usepackage{mathtools}
\usepackage{relsize}
\usepackage{aas_macros}

\makeatletter % some generic helpers
\newcommand{\LCASES}[1]{$\m@th\displaystyle{#1}$\hfil}
\newcommand{\CCASES}[1]{\hfil$\m@th\displaystyle{#1}$\hfil}
\newcommand{\RCASES}[1]{\hfil$\m@th\displaystyle{#1}$}
\makeatother

\newcommand{\lambdacen}{\lambda_{\rm cen}}
\newcommand{\rmis}{r_\mathrm{mis}}
\newcommand{\fmis}{f_\mathrm{mis}}
\newcommand{\lambdamis}{\lambda_\mathrm{mis}}
\newcommand{\hMpc}{h^{-1}\ {\rm Mpc}}
\newcommand{\Omegab}{\Omega_{\rm b}}

\newcases{ecases}{\quad}{\CCASES{##}}{\LCASES{##}}{\lbrace}{.}
\newcases{ecases*}{\quad}{\CCASES{##}}{{##}\hfil}{\lbrace}{.}

\def\beq{\begin{eqnarray}}
\def\eeq{\end{eqnarray}}

\def\siglogm{\sigma_{\log M}}
\def\Mmin{M_\mathrm{min}}
\def\Msun{M_\odot}

\def\redmapper{redMaPPer}
\def\Rlambda{R_\lambda}

\def\pimax{\Pi_\mathrm{max}}
\def\Ncyl{N_\mathrm{cyl}}

\def\ncen{n_{\rm cen}}

\def\DScen{{\Delta\Sigma_{\rm cen}}}
\def\DSmis{{\Delta\Sigma_{\rm mis}}}
\newcommand{\Sbin}{S_{\rm bin}}
\newcommand{\Omegafid}{\Omega_{\rm fid}}
\newcommand{\Omegam}{\Omega_{\rm m}}
\newcommand{\Rfid}{R_{\rm fid}}
\newcommand{\DSfid}{\Delta\Sigma_{\rm fid}}
\newcommand{\calR}{\mathcal{R}}
\newcommand{\DSobs}{\Delta\Sigma_{\rm obs}}

\defcitealias{DESY1CL_2020_et_al}{DES20CL} % Heidi: making DES20CL hyperlinks

\begin{document}

\preprint{APS/123-QED}

\title{Cosmological Constraints from Dark Energy Survey Year 1 Cluster Lensing and Abundances with Simulation-based Forward-modeling}

\author{Andr\'{e}s N. Salcedo}
\email[]{ansalcedo@arizona.edu}
\affiliation{ Department of Astronomy/Steward Observatory, University of Arizona, 933 North Cherry Avenue, Tucson, AZ 85721-0065, USA }
\affiliation{Department of Physics, University of Arizona, 1118 East Fourth Street, Tucson, AZ 85721, USA.}
\author{Eduardo Rozo}
\affiliation{Department of Physics, University of Arizona, 1118 East Fourth Street, Tucson, AZ 85721, USA.}
\author{Hao-Yi Wu}
\affiliation{Department of Physics, Southern Methodist University, Dallas, TX 75205, USA}
\author{David H. Weinberg}
\affiliation{Center for Cosmology and AstroParticle Physics (CCAPP), the Ohio State University, Columbus, OH 43210, USA}
\affiliation{Department of Astronomy, the Ohio State University, Columbus, OH 43210, USA}
\author{Pranav Chiploonkar}
\affiliation{ Department of Astronomy/Steward Observatory, University of Arizona, 933 North Cherry Avenue, Tucson, AZ 85721-0065, USA }
\author{Chun-Hao To}
\affiliation{Department of Astronomy and Astrophysics, University of Chicago, Chicago IL 60637, USA}
\affiliation{Kavli Institute of Cosmological Physics, University of Chicago, Chicago IL 60637, USA}
\affiliation{NSF-Simons AI Institute for the Sky (SkAI), 172 E. Chestnut St., Chicago IL 60611, USA}
\author{Shulei Cao}
\affiliation{Department of Physics, Southern Methodist University, Dallas, TX 75205, USA}
\author{Eli S. Rykoff}
\affiliation{Kavli Institute for Particle Astrophysics \& Cosmology, P. O. Box 2450, Stanford University, Stanford, CA 94305, USA}
\affiliation{SLAC National Accelerator Laboratory, Menlo Park, CA 94025, USA}
\author{Nicole Marcelina Gountanis}
\affiliation{Center for Cosmology and AstroParticle Physics (CCAPP), the Ohio State University, Columbus, OH 43210, USA}
\affiliation{Department of Astronomy, the Ohio State University, Columbus, OH 43210, USA}
\author{Conghao Zhou}
\affiliation{Physics Department, University of California, Santa Cruz, CA 95064, USA}
\date{\today}% It is always \today, today,

\begin{abstract}
We present a simulation-based forward-modeling framework for cosmological inference from optical galaxy-cluster samples, and apply it to the abundance and weak-lensing signals of DES-Y1 redMaPPer clusters. The model embeds cosmology-dependent optical selection using a counts-in-cylinders approach, while also accounting for cluster miscentering and baryonic feedback in lensing. Applied to DES-Y1, and assuming a flat $\Lambda$CDM cosmology, we obtain $\Omega_m=0.254^{+0.026}_{-0.020}$ and $\sigma_8=0.826^{+0.030}_{-0.034}$, consistent with a broad suite of low-redshift structure measurements, including recent full-shape analyses, the DES/KiDS/HSC 3$\times$2 results, and most cluster-abundance studies. Our results are also consistent with \textit{Planck}, with the difference being significant at $2.58\sigma$. These results establish simulation-based forward-modeling of cluster abundances as a promising new tool for precision cosmology with Stage~IV survey data.
\end{abstract}

\maketitle

\section{Introduction}
\label{sec:intro}

Galaxy clusters have long been recognized as a powerful probe of dark energy, provided that cluster selection can be adequately characterized \citep{DarkEnergyTaskForce_Albrecht_2009,Weinberg_PhR_2013}.  In optical data this has been especially difficult: in DES-Y1, the redMaPPer abundance analysis was biased by unmodeled selection effects imprinted on the clusters’ weak-lensing signals \citet{DESY1CL_2020_et_al}.  Further work demonstrated that projection effects in galaxy clusters significantly perturb the clusters' weak lensing profiles \citep{Wu_et_al_2022}, explaining the origin of this bias.  These realizations led to improved modeling in which the impact of projection effects was parameterized based on numerical simulations \citep{To_et_al_2021b,Park_et_al_2023}.  While this approach succeeded in achieving unbiased cosmological analyses, the increased number of parameters degraded the constraining power of the cluster samples. 

In parallel, a new generation of simulation-based inference frameworks has emerged for large-scale structure analyses \citep[e.g.][]{Wibking_et_al_2019,Zhai_et_al_2023,Pellejero-Ibanez_et_al_2024,Miyatake_et_al_2022}. We extend this philosophy to optical cluster cosmology by embedding cosmology-dependent selection and projection directly in a fully simulated inference pipeline. In \citep{Salcedo_et_al_2024b} we introduced a forward-modeling framework that populates halos with galaxies and builds synthetic cluster catalogs to model the clusters’ lensing signal.  We found that unlike in the original DES analysis, the abundance and weak-lensing signal of DES-Y1 redMaPPer clusters are consistent with the \textit{Planck} $\Lambda$CDM best fit. Here we extend that framework to enable full cosmological inference and apply it to the DES-Y1 redMaPPer sample. Because we can reliably model cluster selection at lower richness, we include all clusters with $\lambda\geq10$, a threshold half that of \citet{DESY1CL_2020_et_al}, enabling us to increase the statistical precision of our results.

The layout of the paper is as follows.  In Section~\ref{sec:data}, we summarize the observational and simulation data that underpins our analysis. In Section~\ref{sec:finding} we describe how we populate numerical simulations with galaxies, and provide a brief overview of how we use the resulting galaxy catalogs to characterize the weak lensing signal of redMaPPer galaxy clusters.  Section~\ref{sec:abundance} and \ref{sec:lensing} describe how we model the abundance and weak lensing data from the simulations, while section~\ref{subsec:like} describes our likelihood functions and the emulator design used to interpolate simulation results. Section~\ref{sec:results} presents our results, while section~\ref{sec:conclusions} summarizes our results and presents our conclusions.

\section{Data}
\label{sec:data}

\subsection{DES Data}

We analyze measurements of the lensing and abundance of 24,616  clusters identified by the red-sequence Matched-filter Probabilistic Percolation algorithm \citep[redMaPPer;][]{Rykoff_et_al_2014} in the $1321 + 116 \, \mathrm{deg}^2$ of DES-Y1 imaging data \cite{Drlica-Wagner_et_al_DESY1_2018}.  The redMaPPer algorithm identifies galaxy clusters as overdensities of red-sequence galaxies, using a matched filter approach to estimate the membership probability of each red-sequence galaxy within an empirically defined cluster radius. 

Our cluster sample is separated into three redshift bins with bin edges at $z=[0.2,0.35,0.50,0.65]$. Each redshift bin is further separated into six bins of richness with bin edges at $\lambda=[10,14,20,30,45,60,\infty]$. Here, $\lambda$ is the number of optical galaxies associated with a given cluster. We note this choice contrasts with the existing cluster literature utilizing redMaPPer, which is limited to $\lambda > 20$ in order to maintain purity of the resulting sample. Our framework 
obviates the need for a sample with high purity because non-cluster contaminants are self-consistently included in our forward-modeling, enabling us to exploit the statistics at low richness.

For each of our richness and redshift bins, cluster shear profiles were measured using the DES-Y1 {\sc{Metacalibration}} shape catalog \cite{Zuntz_et_al_DESY1_Metacal_2018} and the BPZ photometric redshift catalog \cite{Hoyle_et_al_DESY1_BPZ_2018}.  The profiles were boost factor corrected using the results of \cite{Varga_et_al_DESY1_2019}. The full details of this measurement are found in \cite{McClintock_DES_2019}.

\subsection{Simulation Data}
\label{subsec:sims}

Our analysis forward-models cluster selection by populating dark matter halos in simulations using a Halo Occupation Distribution (HOD) framework. Because the \redmapper\ cluster finding algorithm counts galaxies down to a luminosity threshold of $0.2L_*$ \citep{Rykoff_et_al_2014}, the simulations used in our analysis must resolve halos down to $M_{\rm 200m}\gtrsim 5\times 10^{11}\ h^{-1}\ M_\odot$ \citep{Salcedo_et_al_2024b}.  Assuming these halos are resolved with $\approx 100$ particles, the particle mass must be lower than $5\times 10^9\ h^{-1}\ M_\odot$.  At the same time, accurate modeling of high-mass clusters requires a large simulation volume, and therefore very large numbers of particles. We are able to satisfy these twin constraints using the AbacusSummit simulations suite\footnote{https://abacussummit.readthedocs.io/en/latest/index.html} \citep{Maksimova_Summit_et_al_2021} thanks to their unique combination of volume ($L_\mathrm{box} = 2.0 \, h^{-1} \, \mathrm{Gpc}$) and mass resolution ($N_\mathrm{part} = 6192^3$ particles of mass $M_\mathrm{part} \sim 2 \times 10^9 \, h^{-1} \, \Msun$).

The simulations are run with the {\sc{abacus}} \citep{Metchnik_2009, Garrison_et_al_2018, Garrison_et_al_2019, Garrison_et_al_2021} cosmological N-body code, which uses GPUs and novel computational techniques to achieve high speed and accuracy. The spline force softening length is $\epsilon_g = 7.2 \, h^{-1} \, \mathrm{kpc}$. Our analysis relies on the 52 emulator grid simulations sampling from a parameter space based on a broadened \textit{Planck} posterior \citep{Planck_DR18_2020}.  The samples further impose the requirement that the CMB acoustic scale is fixed at $100 \times \theta_* = 1.041533$.\footnote{Defined by $\theta_* = r_* / D_{\rm M}(z_*)$, where $r_*$ is the comoving sound horizon at recombination and $D_M(z_*)$ is the comoving angular diameter distance to the last scattering surface at redshift $z_*$.} We augment these 52 simulations with an additional simulation at the \textit{Planck} cosmology \citep{Planck_DR18_2020}. We note that all simulations used the same seed, which improves interpolation accuracy across simulations but means our model predictions suffer from the cosmic variance associated with the volume of the AbacusSummit simulations. However, this volume is several times larger than the volume of any of the DES-Y1 redshift bins.  Moreover, the dominant source of statistical uncertainty in the measurements is not cosmic variance but rather shape noise.

Our analysis uses simulation redshift snapshots at $z = 0.3$, $z = 0.4$, and $z = 0.5$. Halos are identified from particle snapshots using the {\sc{CompaSO}} halo finder \citep{Hadzhiyska_COMPASO_et_al_2022}. We use the ``cleaned'' {\sc{CompaSO}} halo catalogs and adopt as the halo center for each of our distinct halos the center-of-mass of their most massive embedded subhalo, as recommended in \citet{Hadzhiyska_COMPASO_et_al_2022}.

\begin{figure}
\centering \includegraphics[width=0.45\textwidth]{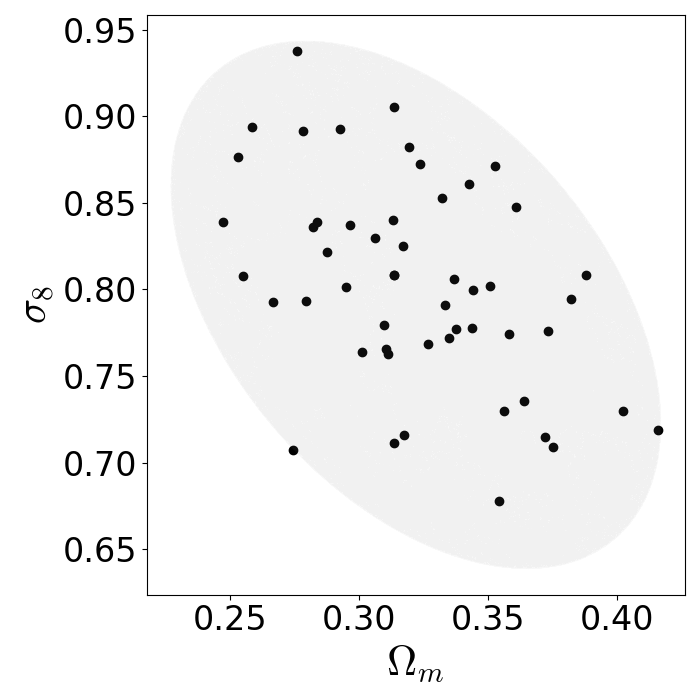}
%\vspace{-0.5cm}
    \caption{The AbacusSummit cosmology sampling in $\Omegam$ and $\sigma_8$ used to train our emulators along with the projection of the AbacusSummit prior ellipsoid in the $\Omegam${-}$\sigma_8$ plane. This ellipsoid is meant to represent a conservative (roughly 6--8$\sigma$) constraint on cosmological parameters from CMB+LSS data circa 2021 \citep{Maksimova_Summit_et_al_2021}.  Our work considers two analyses, one that adopts a flat prior within this ellipsoid, which for brevity we refer to as the ``AbacusSummit'' prior, and one that does \textit{not} adopt this prior.
    } 
\label{fig:summit-samp}
\end{figure}

While the AbacusSummit suite is well-suited to our purposes, it inevitably imposes some external priors on our cluster analysis. First, all of the simulation cosmologies have a fixed $\theta_*$. Second, and more importantly, the range of cosmologies sampled in the AbacusSummit suite is relatively limited in $\Omegam$ and $\sigma_8$, as illustrated in  Figure~\ref{fig:summit-samp}. 
These cosmologies were selected to lie on a multi-dimensional ellipsoid, whose location and extent was meant to comfortably encompass the allowed parameter space from the combination of CMB and large-scale structure data \citep{Maksimova_Summit_et_al_2021}.  The ellipsoid spans between $6\sigma$ and $8\sigma$ fluctuations in cosmological parameters, except for the parameter $\sigma_8$, whose range was further extended by $\pm 0.06$.  

We estimate the ellipsoid enclosing the simulated AbacusSummit cosmologies, and consider two different analyses.  The first analysis adopts a flat prior within the interior of the AbacusSummit ellipsoid and sets the prior to zero outside of this ellipsoid, which is meant to represent a conservative characterization of cosmological constraints at the time the AbacusSummit suite was generated ($\sim 2021$).  For simplicity, we will refer to this conservative CMB+LSS prior simply as the ``AbacusSummit prior.''  By adopting this prior we can ensure that our simulation-based model predictions are never extrapolated.  However, this choice of prior is unfortunately informative: our cluster-based posteriors are clearly truncated by this ellipsoid.  Consequently, we also consider an analysis in which we do \textit{not} include the AbacusSummit prior, enabling us to recover cosmological constraints from cluster abundances alone at the expense of a small amount of model extrapolation. We have no reason to expect a sharp change in model behavior at the boundary of this ellipsoid, so we consider this second case to be a better representation of the constraints from DES-Y1 clusters despite the need for a small amount of model extrapolation.

\section{Synthetic Catalog Generation}
\label{sec:finding}

Our analysis relies on generating synthetic cluster catalogs that mirror redMaPPer in our simulations, and using these to characterize the weak lensing signal of the clusters.  This section gives a general overview of this methodology, with further details provided in Sections~\ref{sec:abundance} and~\ref{sec:lensing}.

\subsection{Halo occupation distribution modeling}
\label{subsec:HOD}

We populate simulated halos with galaxies according to a Halo Occupation Distribution framework \citep[e.g.][]{Benson_et_al_2000, Berlind_2002, vdBosch_et_al_2003b, Zheng_et_al_2005, Zehavi_et_al_2011}. Following standard practice \citep{Zheng_et_al_2005}, we separate the galaxies into satellites and centrals and parameterize their respective mean occupations as
\begin{align}
\langle N_\mathrm{cen} | M \rangle &= \frac{1}{2} \left[ 1 + \mathrm{erf} \left( \frac{\log M - \log \Mmin}{\siglogm} \right) \right], \label{eq:HOD_cen} \\ 
\langle N_\mathrm{sat} | M \rangle &= \langle N_\mathrm{cen} | M \rangle \left( \frac{M}{M_1} \right)^\alpha, \label{eq:HOD_sat}
\end{align}
where the parameter $\Mmin$ represents the characteristic halo mass required to host a central ($\langle N_\mathrm{cen} | \Mmin \rangle = 0.5$), the parameter $\siglogm$ determines the sharpness of the transition from $\langle N_\mathrm{cen} \rangle = 0.0$ to $\langle N_\mathrm{cen} \rangle = 1.0$, $M_1$ is the characteristic halo mass required to host a satellite ($\langle N_\mathrm{sat} | M_1 \rangle = 1.0$), and $\alpha$ is the slope of the satellite-occupation power law. In what follows we will also consider $M_{20}$ as a parameter. $M_{20}$ is the mass at which $\langle N_\mathrm{sat} | M_{20} \rangle = 20$, and is therefore a better representation of the characteristic mass of a galaxy cluster.

Equations~\ref{eq:HOD_cen} and \ref{eq:HOD_sat} give only the mean of the HOD.  The actual number of centrals placed into a given halo is either zero or one, and is determined randomly from the mean central occupation. The number of satellites placed into a halo is drawn randomly from a Poisson distribution with mean $\langle N_\mathrm{sat} | M \rangle.$ We explicitly verify in Appendix~\ref{app:robustness} that allowing for super-Poisson variance in the number of satellites has no impact on our results.  We use the same random number seed when populating every simulation to improve interpolation accuracy across simulations. 

Centrals are placed at the center of their host halo, while satellites are distributed according to a Navarro-Frenk-White profile \citep[NFW;][]{NFW_1997}, parameterized by halo concentrations assigned using the relations provided by \citet{Correa_2015}. These parameters are summarized in Table \ref{tab:param}.  Appendix~\ref{app:robustness} demonstrates our results are insensitive to changing the concentration--mass relations of the galaxies by $\pm 25\%$.

\begin{table*}[]
    \renewcommand*{\arraystretch}{1.1}
    \centering
    \caption{Summary of parameters used in our likelihood analysis. Details of the AbacusSummit cosmology emulation parameter space can be found in \citet{Maksimova_Summit_et_al_2021}. We Latin-hypercube sample our HOD and BCM parameters uniformly in the ranges indicated. Fixed values of the width of our assumed redshift kernel are taken from measurements of \citet{DESY1CL_2020_et_al} using the method of \citet{Costanzi_et_al_2019a}. Priors on miscentering parameters are taken from \citet{Zhang_DES_et_al_2019}.
    }
    \begin{tabular}{cccl}
    \hline 
          Parameter & Emulation Range & Fixed Value or Prior & Description \\
    \hline 
    $\Omegam$ & See $\S$~\ref{subsec:sims} & See $\S$~\ref{subsec:sims} & Cosmological matter density.\\
    $\sigma_8$ & See $\S$~\ref{subsec:sims} & See $\S$~\ref{subsec:sims} & Amplitude of matter clustering. \\
    $\Omegab$ & See $\S$~\ref{subsec:sims} & See $\S$~\ref{subsec:like} & Cosmological baryon density. \\
    \hline
    $n_s$ & See $\S$~\ref{subsec:sims} & 0.9649 & Scalar spectral index. \\
    $w_0$ & See $\S$~\ref{subsec:sims} & -1.0 & Dark energy equation of state normalization. \\
    $w_a$ & See $\S$~\ref{subsec:sims} & 0.0 & Dark energy equation of state slope. \\
    $\alpha_s$ & See $\S$~\ref{subsec:sims} & 0.0 & Running of the spectral tilt. \\
    $N_\mathrm{ur}$ & See $\S$~\ref{subsec:sims} & 2.0328 & Effective number of ultra-relativistic species today. \\
    \hline
    $\mathcal{A}_{m,1}$ & - & $\mathcal{N}(1.021, 0.025)$ & Shear+Photo-$z$ calibration, first redshift bin, $z\in[0.20, 0.35)$ \\
    $\mathcal{A}_{m,2}$ & - & $\mathcal{N}(1.014, 0.024)$ & Shear+Photo-$z$ calibration, second redshift bin, $z\in[0.35, 0.50)$ \\
    $\mathcal{A}_{m,3}$ & - & $\mathcal{N}(1.016, 0.025)$ & Shear+Photo-$z$ calibration, third redshift bin, $z\in[0.50, 0.65)$ \\
    \hline
    $\sigma_{\log M}(0.3)$ & [0.01, 0.60] & $\mathcal{U}(0.01, 0.60)$ & Width of the central occupation at $z = 0.3$. \\
    $\log M_{\mathrm{min}}(0.3)$ & $[12.0, 13.0]$ & $\mathcal{U}(12.0, 13.0)$ & Minimum halo mass to host a central at $z = 0.3$. \\
    $\log M_{20}(0.3)$ & $[13.8, 15.0]$  & $\mathcal{U}(13.8, 15.0)$ & Characteristic mass to host 20 satellites at $z = 0.3$. \\
    $\alpha(0.3)$ & $[0.7, 2.0]$ & $\mathcal{U}(0.7, 2.0)$ & Slope of halo satellite occupation at $z = 0.3$. \\
    \hline
    $\sigma_{\log M}(0.5)$ & [0.01, 0.60] & $\mathcal{U}(0.01, 0.60)$ & Width of the central occupation at $z = 0.5$. \\
    $\log M_{\mathrm{min}}(0.5)$ & $[12.0, 13.0]$ & $\mathcal{U}(12.0, 13.0)$ & Minimum halo mass to host a central at $z = 0.5$. \\
    $\log M_{20}(0.5)$ & $[13.8, 15.0]$  & $\mathcal{U}(13.8, 15.0)$ & Characteristic mass to host 20 satellites at $z = 0.5$. \\
    $\alpha(0.5)$ & $[0.7, 2.0]$ & $\mathcal{U}(0.7, 2.0)$ & Slope of halo satellite occupation at $z = 0.5$. \\
    \hline
    $\sigma_z(0.3)$ & - & 0.065 & Cylinder projection depth at $z = 0.3$. \\
    $\sigma_z(0.4)$ & - & 0.098 & Cylinder projection depth at $z = 0.4$. \\
    $\sigma_z(0.5)$ & - & 0.106 & Cylinder projection depth at $z = 0.5$. \\
    \hline
    $\tau$ & $[0.01, 0.45]$ & $\mathcal{N}(0.166, 0.07)$ & Characteristic scale of miscentering in units of $R_\lambda$.\\
    $f_\mathrm{mis}$ & - & $\mathcal{N}(0.165, 0.09)$ & Fraction of miscentered clusters.\\
    $x_0$ & - & 1.66 & Scale associated with the impact of miscentering on richness. \\
    $a$ & - & 0.26 & Normalization of scatter in miscentered richness correction. \\
    $b$ & - & 1.43 &  Dependence of scatter on miscentering offset. \\
    \hline
    $B$ & $[-2.0, 0.0]$ & $\mathcal{U}(-2.0, 0.0)$ & Baryonification variable (see sec.~\ref{sec:baryons}). \\
    \hline
    \end{tabular}
    \label{tab:param}
\end{table*}

\subsection{Cluster selection with counts in cylinders}
\label{subsec:CIC}

We utilize the counts-in-cylinders approach of \cite{Salcedo_et_al_2024b} \citep[also see][]{Sunayama_et_al_2020, Wu_et_al_2022, Sunayama_et_al_2023, Zeng_et_al_2023} by weighting each galaxy according to its line-of-sight distance from the central halo as per the projection model of \cite{Costanzi_et_al_2019a}:
\beq
    w(d_\mathrm{los}, d_{\rm cyl}) =
    \begin{ecases*}
    1 - \left(\frac{d_\mathrm{los}}{d_{\rm cyl}}\right)^2 & if $|d_\mathrm{los}| < d_{\rm cyl}$, \\
    0 & otherwise.
    \end{ecases*}
\label{eq:cyl_weight}
\eeq
\citet{DESY1CL_2020_et_al} measures the width of this projection kernel in redshift $\sigma_z$ for DES-Y1 clusters using the method of \citet{Costanzi_et_al_2019a}. This measurement of $\sigma_z$ is therefore independent of cosmology. In practice we adopt these measurements of $\sigma_z$ (see Table~\ref{tab:param}) and convert them to the appropriate $d_{\rm cyl}$ given the cosmology of our simulation when applying our mock cluster finder.

We define $N_{\rm cyl}$ as the sum of the galaxy weights of all galaxies within the cylinder, and within a projected aperture
\begin{align}
\Rlambda &= \left( \frac{\Ncyl}{100} \right)^{0.2} \, \left[ h^{-1} \,  \mathrm{physical} \, \mathrm{Mpc} \right]
\end{align}
that mimics that of the \redmapper\ algorithm. 

To generate a cluster catalog, we rank-order halos according to mass.  We then calculate $\Ncyl$ for the most massive halo, and remove all galaxies within the cylinder from consideration for all subsequent halos.  We then move on to the next most massive halo, and iterate until we run out of halos of mass above $10^{11.0} \, h^{-1} \, \Msun$.  The result is a cylindrical cluster catalog in which every cluster is perfectly centered on the most massive halo in the cylinder.

In addition to the HOD parameters, our cylindrical cluster catalogs depend on a variety of parameters and assumptions that we do not vary, specifically: 1) the cylinder depth $d_{\rm cyl}$; 2) the concentration--mass relation used to populate the galaxies in the simulation; 3) the details of the percolation algorithm; and 4) the fact that the HOD scatter is assumed to be Poisson.  In Appendix~\ref{app:robustness}, we demonstrate that our results are insensitive to these details at the statistical precision of DES Y1 data.

\subsection{Model Predictions from Cylindrical Counts}

We have described how to generate synthetic cluster catalogs with a richness measure $N_{\rm cyl}$ that is similar but not identical to the \redmapper\ richness $\lambda$.  To be able to relate our simulations to data we need to determine what the $\lambda$--$\Ncyl$ relation is. Specifically, our analysis proceeds as a three-step process:
\begin{itemize}
    \item \textbf{Step 1:} For each simulation, we use \textit{abundance matching} to determine the $\lambda$--$\Ncyl$ relation that enables us to exactly reproduce the observed abundances in the data.  The resulting $\lambda$--$\Ncyl$ relation depends on cosmological, HOD, projection, and miscentering parameters. 
    \item \textbf{Step 2:} For each halo in a simulation, we use the $\lambda$--$\Ncyl$ relation derived in Step 1 to assign a \redmapper\ richness $\lambda$ to simulated halos, and use the simulation to predict the corresponding weak lensing signal $\Delta\Sigma(R )$ for clusters in a given richness and redshift bin.
    \item \textbf{Step 3:} With our predictions for $\Delta\Sigma(R|\lambda)$ in hand, we compare our predictions to the DES weak lensing data at the likelihood level to arrive at the model posteriors.
\end{itemize}
The next three sections go through each of these three steps in detail.  For readers not interested in the precise technical details of our work, we recommend skipping sections~\ref{sec:abundance} and \ref{sec:lensing}, and reading instead the user-friendly summary of these sections presented in section~\ref{subsec:model_summary}.

\section{Step 1: Abundance Matching}
\label{sec:abundance}

\subsection{The Abundance Matching Framework}

In our algorithm so far, the cylindrical clusters are all perfectly centered. If all \redmapper\ clusters were perfectly centered, we could readily recover the $\lambda$--$\Ncyl$ relation via abundance matching.  That is, we set
\beq
n_{\rm halos}(N_{\rm cyl}) = \ncen(\lambdacen),
\label{eq:AB_matching}
\eeq
where $n_{\rm halos}(\Ncyl)$ is the cumulative halo density in the simulation, and $\ncen(\lambdacen)$ is the cumulative cluster density of perfectly centered clusters in the data.  We can then use Equation~\ref{eq:AB_matching} to solve for $\lambdacen(\Ncyl)$. This relation will enable us to assign a \redmapper\ richness to every halo in a simulation in a way that reproduces the DES abundance data.

In practice, not every cluster is perfectly centered. To address this systematic, the necessary first step is to describe our miscentering model.

\subsection{Cluster Miscentering Model}

To account for cluster miscentering we rely on the miscentering model of \citet{Zhang_DES_et_al_2019}. They found the fraction of miscentered clusters $\fmis$ is consistent with being both redshift and richness independent.  Consequently, $\fmis$ is our first miscentering parameter.  The miscentering offset $r_\mathrm{mis}$ of miscentered clusters is drawn from a gamma distribution,
\beq
P(x|\tau) = \frac{x}{\tau^2} \mathrm{exp}\left( - \frac{x}{\tau} \right),
\eeq
where $x = r_\mathrm{mis} / R_\lambda$, and the parameter $\tau$ represents a characteristic offset that is richness and redshift independent.  This is the second miscentering parameter in the model.  Finally, the richness $\lambdamis$  of a miscentered cluster with an offset $\rmis$ is given by 
\beq
\frac{\lambdamis}{\lambda_\mathrm{cen}} \sim \mathcal{N}(\bar{y}(x), \sigma_y(x)). \label{eq:lam_mis}
\eeq
The mean and scatter of the richness miscentering correction are given by
\begin{align}
\bar{y}(x) &= \mathrm{exp}(-x^2/x_0^2), \\
\sigma_y(x) &= a \times \arctan(b x),
\end{align}
where again $x = \rmis / R_\lambda$.  This probability distribution depends on three parameters, $x_0$, $a$, and $b$, for which we adopt the best-fitting values from the DES-Y1 analysis in \citet{Zhang_DES_et_al_2019},  $x_0=1.66 \pm 0.06$, $a = 0.26 \pm 0.04$, and $b = 1.43 \pm 0.22$.  Because the miscentering corrections are small, we anticipate these parameters are measured accurately enough to have a negligible impact on our results.  Consequently, we hold these parameters fixed, and verify this expectation a posteriori. 

With this miscentering model in hand, we can test the impact of cluster miscentering on the cluster abundance function.  Specifically, we use our abundance model to randomly miscenter every \redmapper\ cluster, and compare the abundance function $n(\lambda)$ after miscentering to that before miscentering.  The impact on the resulting abundance functions is $\lesssim 0.2\%$, and is therefore entirely negligible for our purposes.  Consequently, we will simply ignore cluster miscentering for the purposes of constructing the $\lambda$--$\Ncyl$ relation via abundance matching.  We still account for the impact of cluster miscentering in the weak lensing profiles (see below).

\subsection{The $\lambda$--$\Ncyl$ Relation and Richness Assignments}

We use abundance matching (Equation~\ref{eq:AB_matching}) to determine $\lambda(\Ncyl)$ for each of our simulations using the empirical abundance function $n(\lambda)$. The resulting $\lambda$--$\Ncyl$ relation depends only on cosmology and HOD parameters, but not on the miscentering parameters. With the $\lambda$--$\Ncyl$ relation in hand, we are finally in a position to assign a richness $\lambda$ to every cylinder.  If a cluster is well centered, we simply set $\lambda=\lambda(\Ncyl)$.  If a cluster is miscentered, then we randomly draw the miscentering offset $\rmis$ from the distribution $P(x|\tau)$, where $x\equiv \rmis/R_\lambda$, and then draw the miscentered richness $\lambdamis$ using the distribution $P(\lambdamis/\lambdacen|x)$.  We then set $\lambda=\lambdamis$.  A key feature of this method is that every halo is assigned both a well-centered richness \textit{and} a miscentered richness.  In the next section, we describe how we use both of these richness values to characterize the lensing signal of the galaxy clusters in our simulations.

\section{Step 2: Lens Modeling}
\label{sec:lensing}

At this point, we have generated training simulations that sample the space of cosmological and nuisance parameters.  Each simulation has a catalog of ``cylindrical clusters'' of richness $\Ncyl$, which we have mapped to a centered and miscentered \redmapper\ equivalent richness $\lambda$ using abundance matching. Before evaluating the lensing signal of these clusters we must first account for a variety of additional complications, specifically: 1) baryonic feedback; 2) the impact of cluster miscentering on their lensing profiles; and 3) the fact that in observations we do not measure matter densities and projected distances, but rather shear and angular separations.  In this section, we describe how we calculate the predicted lensing signal after accounting for each of these effects.

\subsection{Computing Cluster Lensing}

We begin by describing how we compute the projected surface density $\Sigma(R)$ of the galaxy clusters in our simulations.  We use {\sc{corrfunc}} \citep{corrfunc:17} to compute the real-space cluster--matter cross-correlation $\xi_{cm}(R, \pi)$ in 60 equal logarithmically spaced bins of $R$ covering scales $0.01 < R < 125.0 \, h^{-1} \, \mathrm{Mpc}$ and 200 equal linearly spaced line-of-sight distance $\pi$ bins out to $\pimax = 200.0 \, h^{-1} \, \mathrm{Mpc}$. This real-space correlation function is converted to the surface density $\Sigma$,
\beq
\label{eq:sigma}
\Sigma(R) \equiv 2 \rho_m \int_0^{\pimax} d \pi \, \xi_{cm}(R, \pi),
\eeq
which we will convert into the lensing observable $\Delta \Sigma$ after modeling the impact of baryonic feedback and miscentering.

\subsection{Baryonic Feedback Model}
\label{sec:baryons}

Baryonic feedback modifies the mass distribution in the interior of halos.  We account for this effect using the baryonic feedback model of \citet{Schneider_et_al_2019}.  Specifically, we use analytic models to characterize the response of the halo density profile to baryons, and then apply this analytic response function to our simulations in order to ``baryonify'' our model predictions. 

We start with the baryonification framework developed by \citet{Giri_Schneider_2021} (also see \citet{Schneider_et_al_2019}).  This framework modifies the matter distribution of halos by adding a gas component and a central galaxy component; satellite galaxies are assumed to trace the dark matter.  The model further includes the impact of baryons on the dark matter profile through adiabatic contraction.  The end result is a 7-parameter model that allows one to baryonify dark matter only profiles (for a detailed explanation see Appendix \ref{app:baryon}). We will refer to a baryonified profile as BCM, for Baryonic Correction Model.

Given an analytic dark matter only halo profile and a set of baryonification parameters $\theta$, we can use the methodology of \citet{Giri_Schneider_2021} to construct a new baryonified profile $\rho_{\rm BCM}(r|\theta,m)$. Our analytic model for $\rho_{\rm DMO}$ is the standard halo model expression of the halo--mass correlation function,
\begin{equation}
    \rho_{\rm DMO}(r|M) = Mu(r|M) + b(M)\xi_{\rm lin}(r), 
\end{equation}
where $u$ is a unit-normalized NFW profile \citep{nfw:96}, and $b(M)$ is the halo bias, which we evaluate using the \citet{Tinker_et_al_2012} model.  After baryonifying this model, we integrate both the DMO and BCM profiles along the line of sight to arrive at the projected matter density $\Sigma$.  We then define the \it baryon suppression function \rm $S(R|\theta;M)$ via
\beq
    S(R|\theta;M) = \frac{\Sigma_{\rm BCM}(R|\theta;M)}{\Sigma_{\rm DMO}(R|M)}.
\eeq
The baryonic suppression function $S$ depends on both the baryonification parameters $\theta$ and the mass $M$ of the halo. In principle, $S$ can also depend on cosmology, but since we do not know this dependence a priori, we treat the baryonification parameters as nuisance parameters independent of cosmology.

At this point, we seek to perform some amount of dimensionality reduction.  First, we combine the baryonic suppression functions across all masses and baryonification parameters $\theta$, and perform a principal component analysis. We find that just two principal components can account for 85\% of the total variance.  That is, we can approximate the baryonic suppression function by an expression of the form
\begin{equation}
    S(r|\theta;M) = \bar S(r|\theta) + A(\theta,M)e_1(r) + C(\theta,M)e_2(r),
\end{equation}
where $e_1$ and $e_2$ are the first two principal component vectors.  The coefficients $A$ and $C$ depend on the halo mass $M$ and the baryonification parameters $\theta$.  We find that we can accurately model the mass dependence of $A$ and $C$ as polynomials in log mass, i.e.
\begin{align}
    A(\theta,M) & = \sum_{n=0}^{3} A_n(\theta)(\ln M)^n, \\
    C(\theta,M) & = \sum_{n=0}^{2} C_n(\theta)(\ln M)^n.
\end{align}
The reason we use ``$C$'' as a variable rather than ``$B$'' will be made clear momentarily.

When we plot the various $A_n$ and $C_n$ coefficients against each other, we find that they are linearly related.  That is, \textit{we are able to write all coefficients as a deterministic function of any one single coefficient}.  In our case, we have chosen $A_3$. The specific relations between the various coefficients and $A_3$ are collected in Appendix~\ref{app:baryon}.  For our purposes, the key fact is that the function $A(\theta|M)$ depends only on a single parameter $B \equiv A_3(\theta)$, which we will refer to as our \textit{baryonification variable.} 

\begin{figure*}
\centering \includegraphics[width=0.9\textwidth]{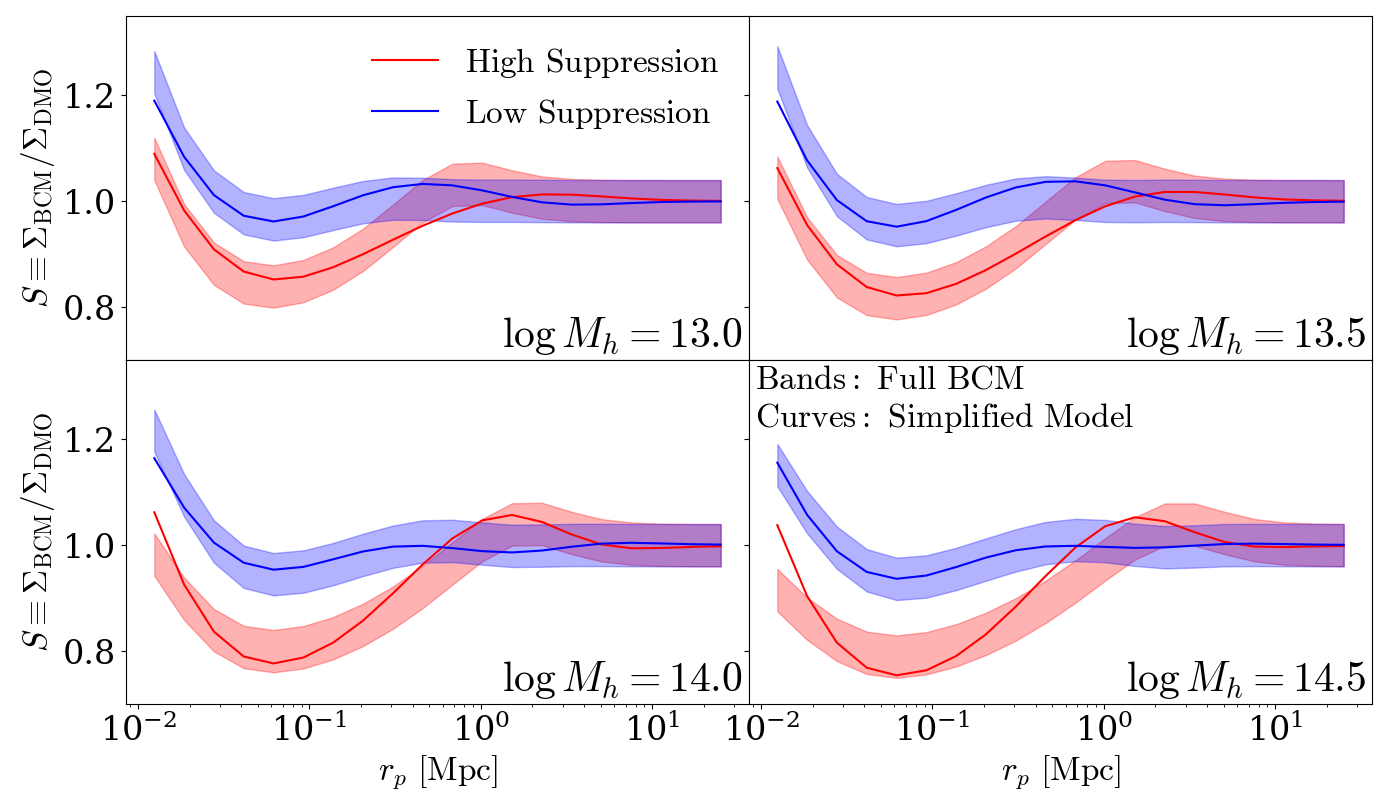}
\vspace{-0.5cm}
    \caption{A comparison of \citet{Giri_Schneider_2021} baryonification models to our one-parameter approximation.  Red and blue correspond to strong and weak baryonic feedback scenarios.  The bands are calculated using the full \citet{Giri_Schneider_2021} models, with the width of the band corresponding to the $\sim4\%$ uncertainty of our emulator. The solid lines are the best-fit models calculated using our dimensionally compressed 1-parameter model, and are fit to all mass bins we consider simultaneously. Here we only plot the fits for four representative masses. The rms difference between the full model and our one-parameter fits is $\sim2\%$.}  
\label{fig:baryons}
\end{figure*}

The performance of this simplified baryonification model is illustrated in Figure~\ref{fig:baryons}, where we compare the baryonic response function as a function of halo mass for random baryonification parameters $\theta$, along with the best fit prediction using our single baryonification variable $B$. Despite some differences, the rms difference between the original \citet{Giri_Schneider_2021} model and our single free-parameter version is $\sim2\%$ when averaged over halos of mass ${\rm log}(M/h^{-1}\ M_\odot) \in [11,15]$ and over the radial range $r\in [0.01,30]h^{-1}\ {\rm Mpc}$.  In light of these results, we replace the full \citet{Giri_Schneider_2021} model by our dimensionally-compressed model that depends only on $B$. 

With this model, we can calculate the baryonic suppression for a given richness and redshift bin.  Specifically, let $M_i$ be the simulated halos within a given richness and redshift bin.  The binned-average baryonic suppression function $\Sbin$ takes the form
\beq
\Sbin(R|B) = \frac{ \sum_{i\in {\rm bin}} \Sigma_{\mathrm{DMO}}(R|M_i)S(R|B;M_i)}{\sum_{i\in {\rm bin}} \Sigma_{\mathrm{DMO}}(R|M_i)}. 
\label{eq:Sbin}
\eeq
In this formula, both $\Sigma_\mathrm{DMO}(R|M_i)$ and $S(R|B;M_i)$ are calculated using our analytic models.  Our baryon-corrected model for the density profile of the galaxy clusters in simulations is given by
\beq
\Sigma_{\rm BCM}(R|B) = \Sbin(R|B) \Sigma(R),
\label{eq:baryonification}
\eeq
where $\Sigma$ is the projected matter density profile of the simulated galaxy clusters in the bin (see equation~\ref{eq:sigma}).

There is one aspect of baryonification that is not properly accounted for in our approach.  Namely, our baryonification procedure is applied only to the most massive main halo for each cluster system, and not to halos along the line of sight. Unfortunately, properly accounting for this effect becomes numerically intractable, though we also anticipate this effect will have a negligible impact on our model: small changes to the profiles of the projected halos have little impact on the projected density profile of the main halo.

\subsection{Incorporating Cluster Miscentering}
\label{sec:miscenter}

Section~\ref{sec:baryons} allows us to baryonify the projected matter density of clusters.  We now turn to incorporating cluster miscentering into our model predictions.  If a fraction $\fmis$ of the clusters are miscentered, then the observed lensing profile is
\beq
\Delta \Sigma &= (1 - f_\mathrm{mis}) \Delta \Sigma_\mathrm{cen} + f_\mathrm{mis} \Delta \Sigma_\mathrm{mis}, \label{eq:delsig_weight_sum}
\eeq
where $f_{\rm mis}$ is the fraction of miscentered clusters in a given richness bin, and $\Delta \Sigma_\mathrm{cen}$ and $\Delta \Sigma_\mathrm{mis}$ refer to the lensing of centered and miscentered clusters, respectively. 

Let us focus on evaluating $\Delta \Sigma_\mathrm{cen}$ first.  The lensing signal of well-centered clusters is related to the \textit{baryonified} matter density $\Sigma_{\rm BCM}$ via
\begin{align}
\label{eq:DelSig}
    \Delta \Sigma_{\rm cen} (R) &= \langle \Sigma(< R) \rangle - \Sigma(R), \nonumber \\
    &=  \frac{2}{r^2_p} \int_0^{R} dr'\ r' \Sigma_{\rm BCM} - \Sigma_{\rm BCM} (R).
\end{align}
The baryonified simulation density profile $\Sigma_{\rm BCM}$ is constructed using all galaxy clusters whose \textit{well-centered} richness $\lambda_{\rm cen}$ falls within the richness bin of interest.

Our model for $\Delta\Sigma_{\rm mis}$ is more complex.  First, we select all simulated clusters whose \textit{miscentered} richness falls within the richness bin of interest.  We then use the simulation particles to evaluate the lensing profile $\Sigma_{\rm DMO}(R)$ \textit{about the true halo centers}.  This is necessary in order to properly account for the impact of baryonification (baryonification and miscentering do not commute).  We evaluate the baryonified profile $\Sigma_{\rm BCM}(R)$ of the galaxy clusters, and then further correct them to account for the impact of cluster miscentering via
\beq
\Sigma_\mathrm{mis}(R | \lambda, z) = \int d r_\mathrm{mis} \, P(r_\mathrm{mis} | \bar \lambda) \Sigma_\mathrm{mis}(R | r_\mathrm{mis}, \lambda, z),\label{eq:mis_dist}
\eeq
where we ignore the mild richness dependence of $P(\rmis)$ with a bin, evaluating the offset distribution at the bin's mean richness. We have demonstrated that this assumption has a negligible impact on the resulting lensing model predictions in Appendix~\ref{app:robustness}. In the above expression, $\Sigma(R|r_{\rm mis})$ is given by 
\beq
\Sigma_\mathrm{mis} %(R | r_\mathrm{mis}) 
    = \int_0^{2 \pi} \frac{d \theta}{ 2 \pi} \, \Sigma_{\mathrm{BCM}} \left( \sqrt{R^2 + r_\mathrm{mis}^2 - 2 R r_\mathrm{mis} \cos \theta } \right). 
\label{eq:mis_one_radii}
\eeq
From here, the miscentered weak lensing profile $\Delta\Sigma_{\rm mis}$ is obtained by using Equation~\ref{eq:DelSig}, only now we replace $\Sigma_{\rm BCM}$ by the miscentered and baryonified profile $\Sigma_{\rm mis}$ calculated using equation~\ref{eq:mis_dist}.  With both $\Delta\Sigma_{\rm cen}$ and $\Delta\Sigma_{\rm mis}$ in hand, we can plug these profiles into Equation~\ref{eq:delsig_weight_sum} to obtain the weak lensing profile of well-centered and miscentered galaxy clusters for each richness and redshift bin in each of our training simulations.

Note that the lensing profiles are baryonified prior to miscentering, as they should be.  Moreover, cluster selection for centered and miscentered clusters is done consistently: clusters that contributed to the well-centered lensing profile of a given bin must have a well-centered richness in that bin.  Likewise, clusters that contributed to the miscentered lensing profile of a given bin must have a miscentered richness in that bin.  This explains why we assigned both a well-centered richness and a miscentered richness to every cluster in our simulations.

\subsection{Cosmology Rescaling}
\label{sec:cosmology_rescaling}

The previous two sections describe how we calculate $\Delta\Sigma(R)$ for clusters in a given richness bin.  In practice, however, we do not measure surface density, but shear.  By the same token, we do not measure projected distance, but rather angular separations.  In the DES analysis, angular separations and shear were transformed into an excess surface density as a function of radius by adopting a fiducial flat $\Lambda$CDM cosmology with $\Omegam=0.3$ and $h=0.7$.  Thus, we must first convert $\Delta\Sigma(R)$ from each simulation into a shear profile $\gamma(\theta)$ using the simulated cosmology, and then convert this shear profile back into an excess surface density profile $\DSfid(\Rfid)$ assuming the DES fiducial cosmology.  

The transverse radial separation conversion is straightforward: in a flat universe, a comoving separation $R$ in a cosmology $\Omega$ corresponds to a transverse angle $\theta = R/\chi(z|\Omega)$, where $\chi(z|\Omega)$ is the comoving distance as a function of redshift in cosmology $\Omega$. This angle in turn corresponds to a transverse separation $\Rfid=\chi(z|\Omegafid) \theta$ in the fiducial cosmology. Thus, the two scales are related via
\beq
R(\Rfid) = \frac{\chi(z|\Omega)}{\chi(z|\Omegafid)}\Rfid.
\eeq

We now turn to the lensing amplitude $\DSfid$ measured at a given scale.  The conversion between shear and excess surface density profile involves the inverse critical surface density
\beq
\Sigma_\mathrm{c}^{-1} = \frac{4 \pi G}{c^2} \frac{D_{\rm l}}{D_{\rm s}} (D_{\rm l} - D_{\rm s}),
\eeq
where $D_{\rm l}$ and $D_{\rm s}$ are the angular diameter distance to lens and source. The estimator used by the DES is
\beq
\hat{\Delta \Sigma} = \frac{\sum w_i \gamma_{t,i} / \langle \Sigma_{c}^{-1} | \Omega_\mathrm{fid} \rangle_i}{\sum w_i},
\eeq
where $i$ indexes all source--lens pairs, and the weights are given by
\beq
w_i \equiv \langle \Sigma_c^{-1} | \Omega_\mathrm{fid}
 \rangle_i^2 / \sigma^2_{\gamma,i},
\eeq
and $\sigma^2_\gamma$ is the shape noise per galaxy. Taking the expectation value in a cosmological model $\Omega$ and assuming a constant shape noise, we find 
\beq
\DSfid(\Rfid) = \calR\Delta\Sigma(R(\Rfid)|\Omega),
\label{eq:scaling}
\eeq
where $\calR$ is a cosmology-dependent rescale factor given by
\beq
\mathcal{R} = \frac{\int d z_l \phi(z_l) \int d z_s \phi(z_s) \Sigma_c^{-1}(z_s, z_l | \Omegafid) \Sigma_c^{-1}(z_s, z_l | \Omega)}{ \int d z_l \phi(z_l) \int d z_s \phi(z_s) \Sigma_c^{-1}(z_s, z_l | \Omegafid)^2}.
\eeq
In this expression, $\phi(z_l)$ and $\phi(z_s)$ are lens and source redshift distributions, and $\Omega$ refers to the cosmology for which we are predicting a $\Delta \Sigma_\mathrm{emu}$.  We calculate this rescaling factor using the DES-Y1 source distributions estimated using the BPZ algorithm \citep{Benitez_BPZ_2000} implemented in \citet{Hoyle_et_al_DESY1_BPZ_2018} to evaluate our model predictions for the observed weak lensing shear signal of the DES galaxy clusters.

\section{Step 3: Likelihood Evaluation}

The previous two sections detailed our simulation-based framework for evaluating the abundance and weak lensing signal of \redmapper\ galaxy clusters using the AbacusSummit simulations. Because those sections were thick in technical detail, we begin this section by offering a brief high-level summary of our model before detailing how we use it to constrain cosmological parameters with the DES \redmapper\ data.

\subsection{Modeling Summary}
\label{subsec:model_summary}

We compute the cluster lensing and abundance for a single set of cosmology, HOD, miscentering, and baryonification parameters at a given redshift as follows:

\begin{enumerate}

\item Populate halos in the AbacusSummit N-body simulation snapshot of that cosmology according to equations \ref{eq:HOD_cen} and \ref{eq:HOD_sat}. 

\item For each halo, we calculate its HOD galaxy count-in-cylinder $\Ncyl$ using the weighting function in equation \ref{eq:cyl_weight}. We then convert $\Ncyl$ into the cluster richness $\lambdacen$ using abundance matching, i.e., equation \ref{eq:AB_matching}. The subscript ``cen'' indicates this richness has not yet been perturbed to account for the impact of cluster miscentering.

\item For each halo, we calculate a miscentered value of the richness $\lambda_\mathrm{mis}$ based on the assumed characteristic miscentering offset parameter $\tau$ and equation \ref{eq:lam_mis}.  In this way, every cluster is assigned both a well-centered and a miscentered richness.

\item To calculate the lensing profile of well-centered clusters, we select all clusters with $\lambdacen$ in the richness bin of interest.  We then use the dark matter particles in the simulation to arrive at the DMO profile $\Delta\Sigma_{\rm DMO}(R)$ in the simulation. We then perturb this profile using our compressed version of the \citet{Schneider_et_al_2019} baryonification model to arrive at the Baryon Corrected Model profile $\Delta\Sigma_{\rm BCM}(R)$ as per equation~\ref{eq:baryonification}.  This defines the lensing signal $\DScen$ of well-centered clusters.

\item To calculate the lensing profile of miscentered clusters, we select all clusters with $\lambdamis$ in the richness bin of interest.  We then use the dark matter particles in the simulation to arrive at the DMO profile $\Delta\Sigma_{\rm DMO}(R)$ in the simulation. Note that to adequately treat baryonification, this lensing profile must be computed about the true halo centers.  After baryonifying the well-centered profile of miscentered clusters, we account for the impact of miscentering in the weak lensing profile as per equation~\ref{eq:mis_dist}.  The result is $\DSmis$, the weak lensing signal of miscentered galaxy clusters in the richness bin of interest.

\item Given $\DScen$ and $\DSmis$, the final weak lensing profile for galaxy clusters in a given richness bin is an appropriately weighted linear combination of the two as per equation~\ref{eq:delsig_weight_sum}.

\item The true lensing observable DES \redmapper\ clusters is not $\Delta\Sigma(R)$, but rather shear and angular separation.  DES converts these into a fiducial projected distance $\Rfid$ and an excess surface density $\DSfid$ by adopting a flat $\Lambda$CDM fiducial cosmology with $\Omegam=0.3$ and $h=0.7$.  We can evaluate the predicted weak lensing signal $\DSfid(\Rfid)$ based on the excess surface density profile $\Delta\Sigma(R)$ via equation~\ref{eq:scaling}.

\end{enumerate}

This process allows us to predict cluster lensing profiles for any set of HOD, projection, miscentering, and baryonification parameters at the cosmology of each of our simulation boxes.

\subsection{Emulator Design}
\label{sec:emu}

In practice, the algorithm detailed in section~\ref{subsec:model_summary} is much too cumbersome to use for parameter inference. Consequently, we use this process to generate a set of model data vectors to train an emulator that enables cosmological parameter inference.

We construct our emulator with a training set generated by a Latin hypercube sampling of the flat priors on our HOD, baryonification parameter $B$, and miscentering parameter $\tau$ (see Table~\ref{tab:param}). We generate 106 such samples, and assign 2 of these parameterizations without replacement to each of our 53 cosmologies. We then use the algorithm summarized in Section~\ref{subsec:model_summary} to compute the predicted lensing signal $\DSfid(\Rfid)$ for each of these simulations. The data vector $\DSfid(\Rfid)$ is defined using the same richness, redshift, and radial bins employed in the real data.  We treat each element in this data vector as independent of the rest, and use Gaussian processes to interpolate across our full parameter space on an element-by-element basis.

To account for redshift evolution across the survey, the emulator for each redshift bin is constructed using a different snapshot from the AbacusSummit simulations.  We use the $z=0.3$ snapshot to build the $z\in[0.2,0.35]$ emulator, the $z=0.4$ snapshot to build the $z\in [0.35,0.5]$ emulator, and the $z=0.5$ snapshot so build the $z\in[0.5,0.65]$ emulator. In \citet{Salcedo_et_al_2024b} we demonstrated that the fact that these snapshots do not exactly coincide with the mean/median redshift of each bin results in negligible systematic uncertainties, even for the case of our highest redshift bin.

To improve the accuracy of our emulator, after running a first analysis using the emulator described above, we sample an additional set of 106 HOD and nuisance parameters from the posteriors, and use these to augment the initial 106 training simulations.

\begin{figure}
\centering \includegraphics[width=0.45\textwidth]{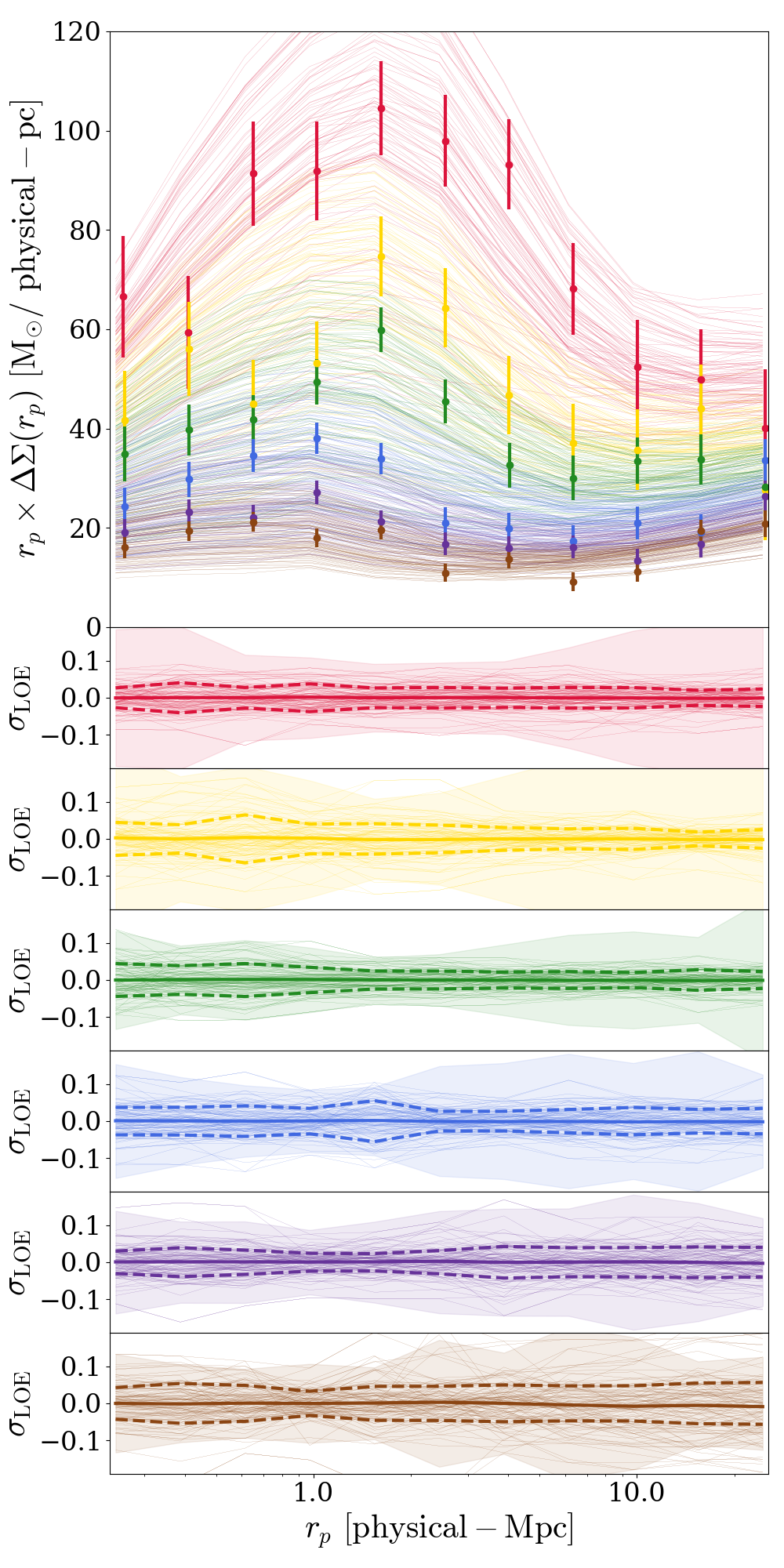}
\vspace{-0.5cm}
    \caption{Lensing profiles in redshift bin $z \in [0.20, 0.35)$ (points with errorbars) compared to simulation predictions for our training data (lines); we emphasize that data is included only for reference and that we are not presenting a fit. Colors correspond to richness bins $\lambda \in [60, \infty)$ (red), $[45, 60)$ (yellow), $[30, 45)$ (green), $[20, 30)$ (blue), $[14, 20)$ (purple), and $[10, 14)$ (brown). The bottom panels show the leave-one-out error for each richness bin compared to the observable covariance (band). The solid and dashed lines correspond to the mean and $1\sigma$ errors, respectively.} 
\label{fig:emu-acc}
\end{figure}

Figure \ref{fig:emu-acc} illustrates the accuracy of our final emulator. The top panel shows our data vector (points with error bars in the lowest redshift bin ($z \in [0.20, 0.35)$) compared to our training set (light lines). Bottom panels show the fractional leave-one-out error (dashed lines) relative to the observable covariance (bands).   Note that our ``leave one out'' test removes one N-body simulation, corresponding to four distinct training samples. The final systematic uncertainty in the emulator profiles is $\lesssim 4\%$, compared to the $10{-}20\%$ errors in the data. This additional uncertainty is included in the final covariance used for inference by adding it in quadrature with the analytically estimated covariance described in Section~\ref{subsec:like}.

\subsection{Gaussian likelihood model}
\label{subsec:like}

Because the abundance data is used to determine the $\lambda$--$\Ncyl$ relation, the data vector in our likelihood is restricted to the observed weak lensing profile $\DSobs(\Rfid)$ in richness and redshift bins.

To constrain our model parameters, we assume a Gaussian likelihood model, $\mathcal{L} \propto e^{- \chi^2 /2}$, where
\beq
\chi^2 = \sum_{i,k,m,n} \Delta(r_{p,m}, z_i, \lambda_k) \left[C^{-1}_{z_i,\lambda_k} \right]_{mn} \Delta(r_{p,n}, z_i, \lambda_k),
\eeq
and $\Delta \equiv \Delta \Sigma_\mathrm{emu} - \Delta \Sigma_\mathrm{obs}$ is the difference between emulated and measured observable cluster lensing, and $C$ is the lensing covariance matrix. We compute these covariance matrices using the Gaussian formalism presented in \citet{Wu_et_al_2019},
\begin{align}
    &C[\Delta\Sigma(r_{p,m}), \Delta \Sigma (r_{p,n})] = \frac{1}{4 \pi f_\mathrm{sky}} \int \frac{k dk}{2 \pi} \hat{J}_2 (k r_{p,m}) \hat{J}_2 (k r_{p,n})  \nonumber \\
    &\times \left[\left(C^{hh}_\ell + \frac{1}{n_h^{2D}}\right)\left( C_\ell^{\Sigma \Sigma} + \langle \Sigma_\mathrm{crit} \rangle^2 \frac{\sigma_\gamma^2}{n_s^{2D}} \right) + \left(C^{h\Sigma}_{\ell}\right)^2\right],
\end{align}
where $f_\mathrm{sky}$ is the DES-Y1 fraction of the sky, $\hat{J}_2$ are radially bin averaged Bessel functions of the second kind, $C^{cc}_\ell$, $C^{\Sigma\Sigma}_\ell$ and $C^{c \Sigma}_\ell$ are the cluster-cluster, projected matter-matter and cluster-projected matter angular power spectra, $n_c^{2D}$ and $n_{s}^{2D}$ are the surface number density of clusters and source galaxies respectively, and $\sigma_\gamma$ is the DES-Y1 shape noise per source galaxy. For each redshift and richness bin we compute relevant cluster power spectra using the \citet{Tinker_et_al_2012} bias of the relevant mean mass calibrated by \citet{McClintock_DES_2019}. The covariance matrix is non-diagonal across radial bins, but different richness bins are assumed to be uncorrelated. This is a reasonable approximation because the errors are shape-noise-dominated. We refer the reader to \citet{Wu_et_al_2019} for further details. We then rescale each of our covariances so that their diagonal elements agree with the variance estimated from jackknife on the data. This rescaling is typically $10-20\%$, but is unusually large for our highest richness and redshift bin, reaching a factor of two. Our results are negligibly impacted by dropping this bin.  For instance, the change in our best fit $S_8$ parameter is $0.1\sigma$, while the error bar changes by only $3\%$.

The model predictions for $\Delta\Sigma$ are taken directly from the emulator described in the previous section.  We emulate the centered and miscentered lensing profiles independently and calculate the final lensing profile via
\beq
\Delta \Sigma_\mathrm{emu} &\equiv \mathcal{A}_m \mathcal{R}  \left[ f_\mathrm{mis} \Delta \Sigma_\mathrm{emu}^\mathrm{mis} + ( 1 - f_\mathrm{mis}) \Delta \Sigma_\mathrm{emu}^\mathrm{cen}\right],
\eeq
where $f_\mathrm{mis}$ is the fraction of miscentered clusters, $\Delta \Sigma^\mathrm{mis}_\mathrm{emu}$ and $\Delta \Sigma^\mathrm{cen}_\mathrm{emu}$ are the miscentered and centered lensing emulator predictions, and $\mathcal{A}_m$ accounts for a possible modulation of the weak lensing signal due to shear calibration and source photometric redshift uncertainties \citep{McClintock_DES_2019}.
We implement linear redshift evolution of our HOD parameters in our likelihood analysis by parameterizing each of our HOD parameters (i.e., $\sigma_{\log M}$, $\log \Mmin$, $\log M_{20}$ and $\alpha$) separately in our lowest and highest cluster redshift bins, and assume a linear evolution in redshift between the two to predict the value of these parameters in the middle redshift bin.

\begin{figure*}
\centering \includegraphics[width=1.0\textwidth]{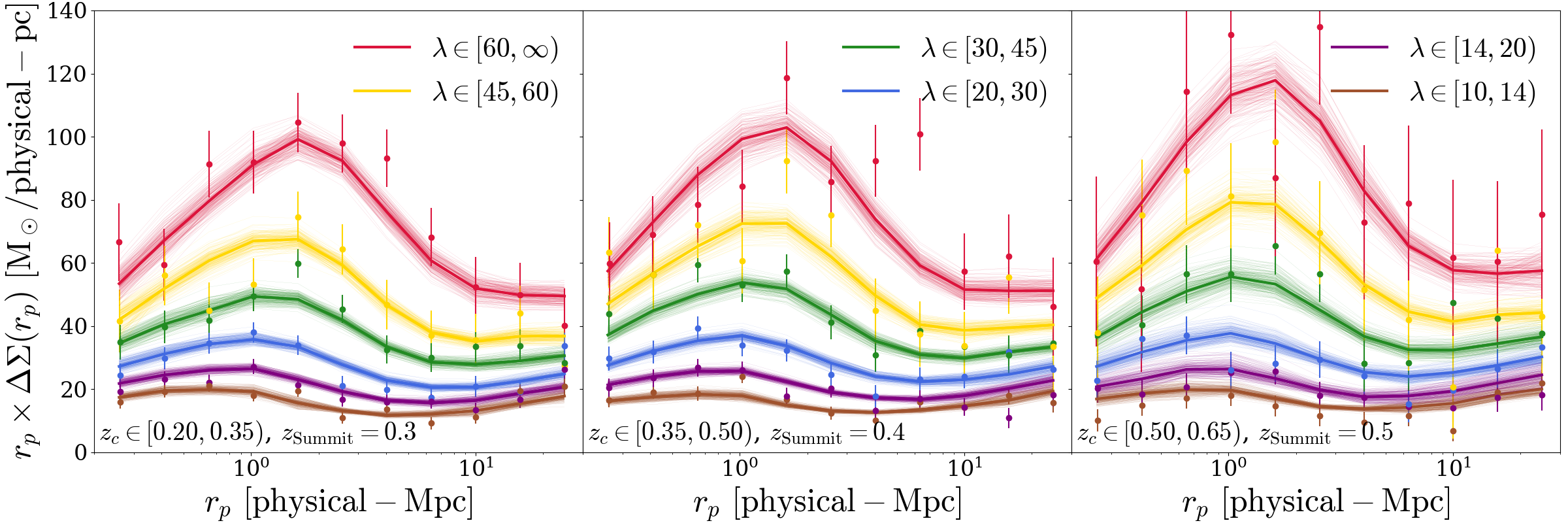}
\caption{Comparison of DES-Y1 cluster weak lensing profiles (point with errorbars) in redshift bins $z\in[0.20, 0.35)$ (left panel), $z\in[0.30, 0.50)$ (middle panel), and $z\in[0.50,0.65)$ (right panel) with those predicted by our fiducial posterior mean model and 200 random samples from our fiducial MCMC chain.  The $\chi^2$ of the best fit model is $\chi^2/\mathrm{dof}=190.51/188$.}
\label{fig:best-fit}
\end{figure*}

Finally we include a prior on the CMB acoustic scale that fixes $100 \times \theta_{\star} = 1.041533$ as is done for the AbacusSummit emulator grid, as well as a conservative Gaussian prior on the baryon density $\omega_{\rm b} = \Omega_{\rm b} h^2$ of $\mathcal{N}(0.02208, 0.00052)$ defined in \citet{DESY1_H0_BAO_DoH} based on Big Bang Nucleosynthesis constraints \citep{Cooke_et_al_2016}. As noted earlier, in some cases we also impose a top-hat prior within the multi-dimensional ellipsoid from which the AbacusSummit simulation cosmologies are drawn. We sample our parameter posteriors using the affine invariant ensemble Markov-Chain-Monte-Carlo (MCMC) sampler implemented in {\sc{emcee}} Python package \cite{Foreman-Mackey_et_al_2016}.

\section{Results}
\label{sec:results}

\subsection{Fiducial results and systematics tests}
\label{subsec:fid_results}

\begin{figure*}
    \centering \includegraphics[width=0.7\textwidth]{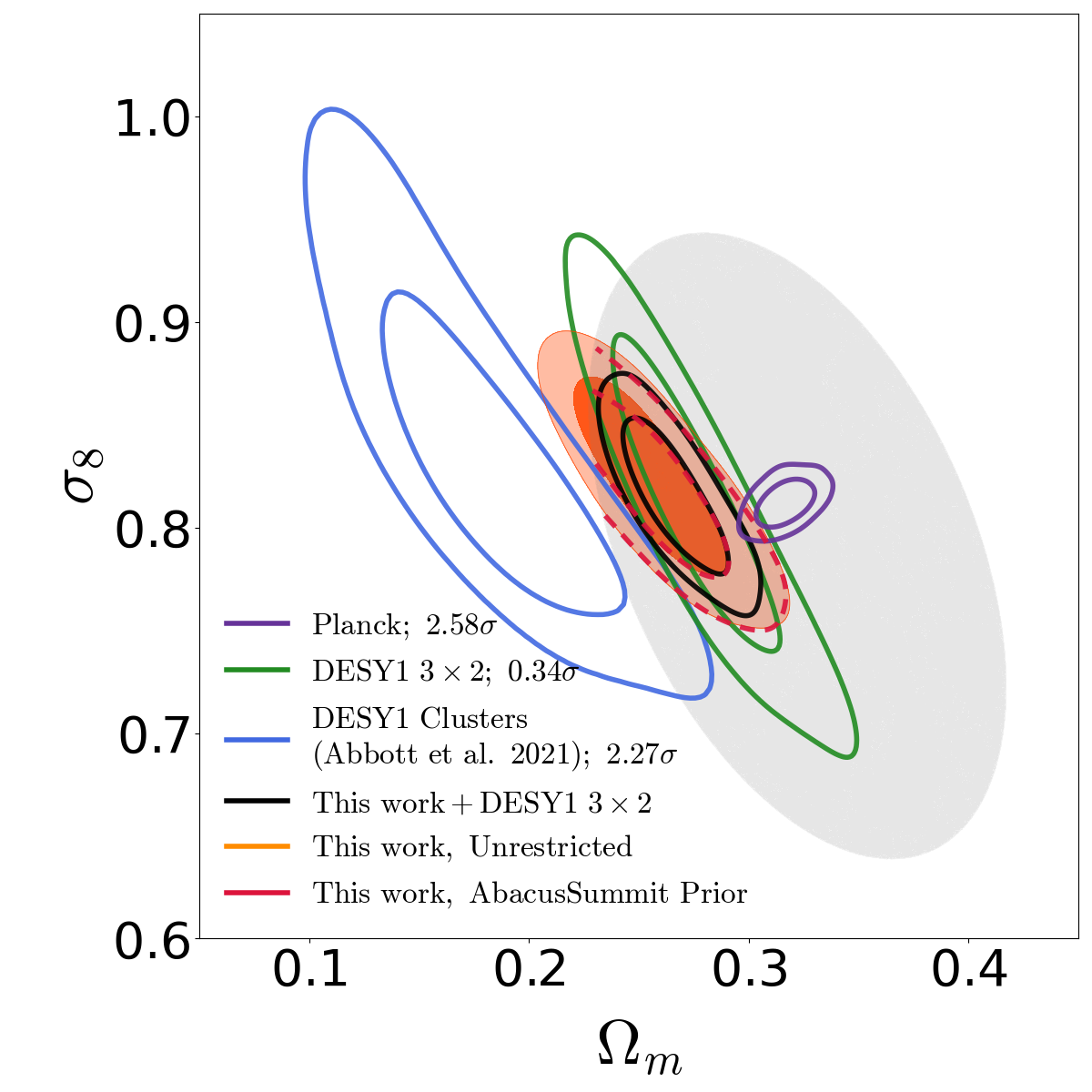}
    \caption{$68\%$ and $95\%$ confidence contours in the $\sigma_8$--$\Omegam$ plane for our analysis with (red dashed) and without (orange filled) the AbacusSummit sampling prior (gray). Tensions with other data sets are quoted with respect to the posterior without the AbacusSummit prior. We compare our results with those from the original DES-Y1 analysis of cluster lensing and abundance \citep[blue;][]{DESY1CL_2020_et_al}, DES-Y1 3$\times$2 \citep[green;][]{DES_3x2pt_2018}, and constraints from {\it{Planck}} CMB measurements \citep[purple;][]{Planck_DR18_2020}. The result of combining our cluster constraints with those from DES-Y1 $3\times2$ is shown in black.}
    \label{fig:fid_constraints}
\end{figure*}

We use the abundance and lensing signal of  DES-Y1 redMaPPer clusters in six richness bins and three redshift bins to derive cosmological constraints.  The sample includes all clusters with richness $\lambda\geq 10$ and spanning the redshift range $z=0.20-0.65$. To model this data we employ the simulation-based framework described in Sections \ref{sec:finding}{-}\ref{sec:lensing} and summarized in Section \ref{subsec:model_summary}. Our Gaussian likelihood model relies on analytically estimated covariance matrices that are rescaled to match Jackknife estimates of the data, as described in detail in \ref{subsec:like}. Our analysis holds the spectral index of the primordial matter fluctuations fixed to $0.9649$, and the effective number of relativistic species today to $N_{\rm ur}=2.0328$, as appropriate for one massive neutrino species of minimal mass.\footnote{This value corresponds to the same early relativistic energy density as the $N_{\rm eff}=3.046$ value for massless neutrinos.}  We do not allow for running of the spectral index, and we fix the dark energy equation of state to $w=-1$.  Our primary goal is to set constraints on $\sigma_8$ and $\Omegam$, which are the two cosmological parameters with the largest impact on cluster weak lensing profiles. For further details, see Table~\ref{tab:param}.

Figure \ref{fig:best-fit} compares the  DES weak lensing data to the weak lensing profiles obtained by sampling our model posterior when imposing the AbacusSummit prior --- i.e., a conservative CMB+LSS prior circa 2021, as described in Sec.~\ref{subsec:sims}. Our best-fit model adequately describes the data with a $\chi^2 / \mathrm{dof} =190.51/188$ and a corresponding probability-to-exceed value of $0.435$. This best-fitting point lies comfortably within the AbacusSummit prior.

Figure \ref{fig:fid_constraints} shows our marginalized constraints with (red) and without (orange) the AbacusSummit prior (gray) in the $\Omegam{-}\sigma_8$ plane. We can see that while the run without the AbacusSummit prior extends past it, thereby requiring some amount of model extrapolation, the posterior volume that requires extrapolation is small: $\approx$10\% of the points in our chain fall outside the AbacusSummit prior.  In Appendix~\ref{app:extrapolation}, we present tests that further suggest that this limited extrapolation does not significantly impact our results. 

Figure~\ref{fig:fid_constraints} also compares our posteriors to those from {\it{Planck}} CMB measurements \citep[purple;][]{Planck_DR18_2020}, the DES-Y1 3$\times$2 \citep[green;][]{DES_3x2pt_2018}, and the original DES-Y1 cluster lensing and abundance analysis \citep[blue;][]{DESY1CL_2020_et_al}. Compared to the original DES-Y1 cluster analysis in \citet{DESY1CL_2020_et_al}, our constraints are significantly shifted to higher $\Omegam$.  This is not simply due to the AbacusSummit prior: as demonstrated in \citet{Salcedo_et_al_2024b}, our simulation framework results in acceptable fits to the lensing data for a \textit{Planck} cosmology, whereas the original model from \citep{DESY1CL_2020_et_al} exhibits a much larger $\chi^2$ ($\Delta\chi^2 =39.91$). Additionally, Figure~\ref{fig:best-fit} shows that our best fit is well within the AbacusSummit sampling region, and that it provides an excellent description of the data.

Following \citet{Lemos_et_al_2021} we calculate the relative $\chi^2$ between two posteriors A and B as
\beq
\chi^2 = (\vec{p}_A - \vec{p}_B)^T (\mathrm{Cov}_A + \mathrm{Cov}_B)^{-1} (\vec{p}_A - \vec{p}_B),
\eeq
where $\vec{p}_A$ and $\vec{p}_B$ are the $(\Omegam, \sigma_8)$ values from posterior A and B respectively, and $\mathrm{Cov}_A$ and $\mathrm{Cov}_B$ are the corresponding $2\times2$ posterior parameter covariance matrices. We then convert this $\chi^2$ into a probability to exceed value and report the corresponding tension in units of $\sigma$.

Using our runs without an AbacusSummit prior, we find our constraints are in $2.58\sigma$ tension with \textit{Planck}, and in $2.86\sigma$ tension with the combined CMB-primaries from \textit{Planck}, ACT, and SPT \citep{Camphuis_et_al_2025}. Note the tension with \textit{Planck} is much reduced from the $5.6\sigma$ tension reported in \citep{DESY1CL_2020_et_al} despite the fact that our results are significantly more precise. Our posteriors are now also comfortably consistent with the DES-Y1 3$\times$2-point analysis.  We compare to DES-Y3 results below, but the juxtaposition here shows that cluster constraints are similar in precision to the $3\times2$ constraints using the same weak lensing data set.

Our constraints on key cosmological parameters when we do \textit{not} impose the AbacusSummit prior are
\begin{align}
\Omegam &= 0.254^{+0.026}_{-0.020},\nonumber \\
\sigma_8 &= 0.826^{+0.030}_{-0.034}, \\
S_8 &= 0.759^{+0.020}_{-0.019}. \nonumber 
\end{align}
These constraints are marginalized over all HOD, miscentering, and baryonification parameters.  If we do impose the AbacusSummit prior we find $S_8 = 0.760^{+0.021}_{-0.018},$ which is indeed nearly identical to the posterior obtained without the prior. For $\Omegam$, the AbacusSummit prior is informative because values below $\sim0.23$ are not allowed. From here onward we concentrate on results without the AbacusSummit prior, as we consider these a more accurate representation of the cluster constraints.

Because our cluster lensing data vector is shape-noise dominated while the DES-Y1 $3\times2$ data vector is sample variance dominated, the two probes are approximately independent. We therefore combine\footnote{This combination was done using the {\sc{CombineHarvesterFlow}} software package \citep{Taylor_CombineHarvesterFlow_et_al_2024}.} the two analyses, finding
\begin{align}
    \Omegam &= 0.266^{+0.017}_{-0.015},\nonumber \\
    \sigma_8 &= 0.815^{+0.026}_{-0.024}, \\
    S_8 &=  0.767^{+0.015}_{-0.014}. \nonumber 
\end{align}
These constraints are shown in Figure~\ref{fig:fid_constraints} as black contours.  They are in $2.99\sigma$ tension with \textit{Planck} and $3.47\sigma$ with the combination of Planck, SPT, and ACT CMB primaries \citep{Camphuis_et_al_2025}.  We note that after combining with DES Y1 3$\times$2-point data, only 0.7\% of the posterior falls outside of the AbacusSummit prior.

\subsection{Robustness Tests}
We assess the robustness of our constraints by splitting our data vector into different disjoint subsamples as described below.  Because the posteriors from these subsets are less constraining, and would therefore require larger extrapolations, our robustness tests are performed with respect to the parameter combination
\begin{equation}
    \Sigma_8\equiv \sigma_8(\Omegam/0.3)^{0.36}
\end{equation}
that is best constrained by the data. 

\begin{figure}
    \centering \includegraphics[width=0.45\textwidth]{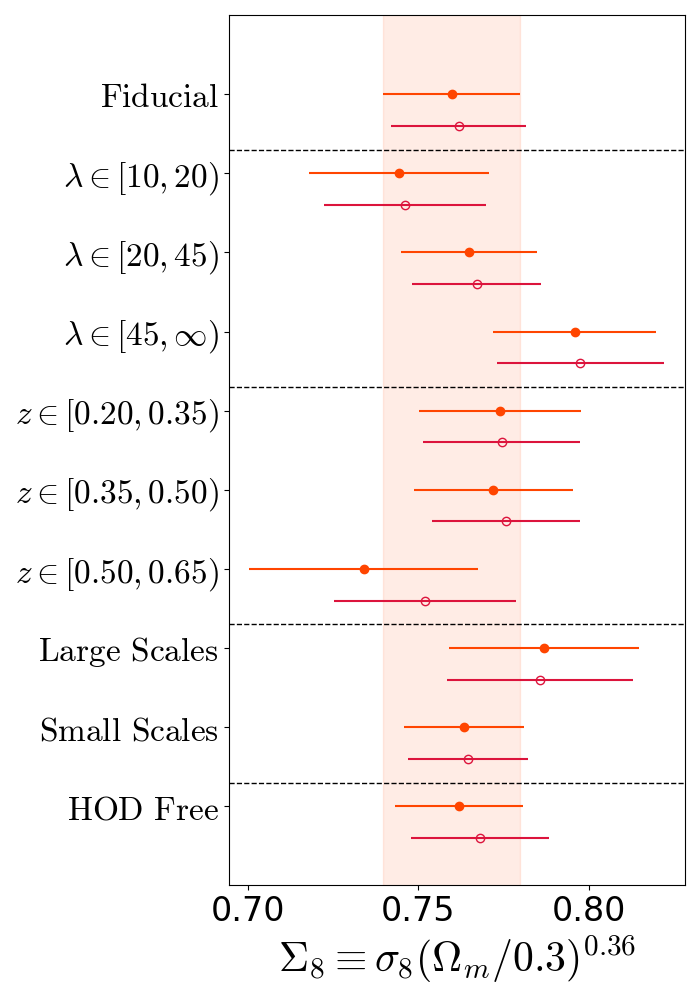}
    \caption{Marginalized constraints on $\Sigma_8 \equiv \sigma_8 \left( \Omegam / 0.3 \right)^{0.36}$, the best constrained combination of $\sigma_8$ and $\Omegam$ for our fiducial unrestricted results shown in Figure~\ref{fig:fid_constraints}, for a variety of analysis scenarios.  In each case, the filled orange point with errorbar shows constraints for the unrestricted case, and the unfilled red point shows constraints when the AbacusSummit sampling space is adopted as an informative prior.}
    \label{fig:robust_tests}
\end{figure}

The data splits we consider are:
\begin{itemize}
    \setlength\itemsep{1pt}
    \item \textit{Richness splits:} We split the six richness bins into three subsets, each with two richness bins, and derive cosmological constraints using those bins only.
    \item \textit{Redshift splits:} We derive cosmological constraints using only one redshift bin at a time.
    \item \textit{Scales splits:} We derive cosmological constraints by restricting our analysis to scales either below or above a radial threshold of $R=2\ \hMpc$.
\end{itemize}
Finally, we also test whether the evolution model impacts our posteriors by treating the HOD parameters in each of the three redshift bins as being independent of one another.  The resulting $\Sigma_8$ posteriors are shown in Figure~\ref{fig:robust_tests} for cases with (orange) and without (red) the AbacusSummit prior imposed.  We see that the $\Sigma_8$ posteriors are consistent across all samples, demonstrating the robustness of our results. One possible source of concern is that $\Sigma_8$ increases monotonically as we go from the low richness bins to the high richness bins.  However, it is not obvious that this trend is significant: the three richness bins are independent random draws, in which case one naively expects a probability of 1/3 for the middle bin to fall between bins 1 and 3.

\begin{figure*}
\centering \includegraphics[width=0.95\textwidth]{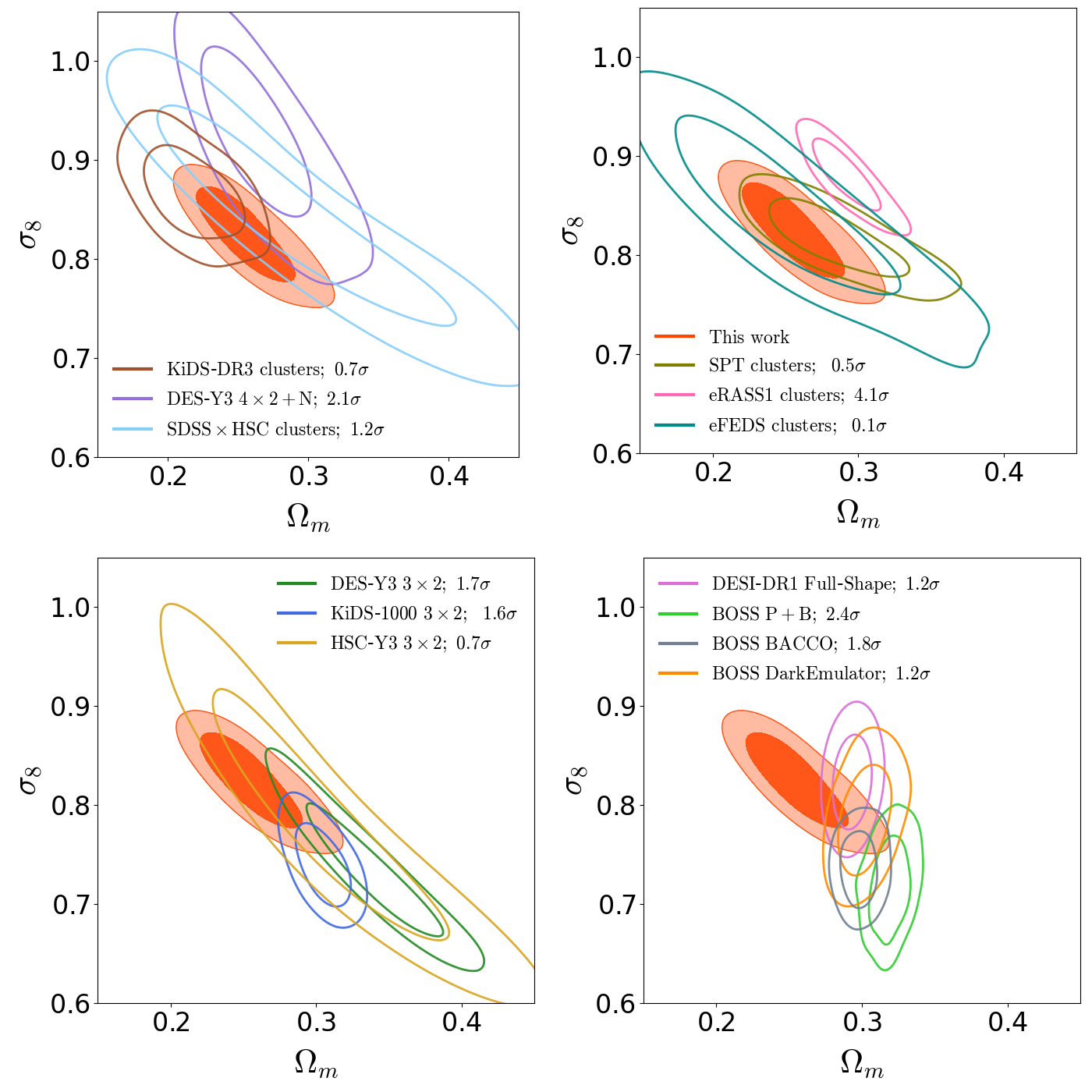}
\caption{68\% and 95\% confidence contours in $\sigma_8$--$\Omegam$ for a variety of different experiments. The legends in each panel quote the tension of our ``unrestricted'' results with each data set, as labeled. {\bf{Top-left:}} Comparison with other Stage-III optical cluster constraints. {\bf{Top right:}} Comparison with X-ray (eROSITA) and SZ (SPT) cluster cosmology constraints. {\bf{Bottom-left:}} Comparison with Stage-III 3$\times$2 constraints. {\bf{Bottom right:}} Comparison with full-shape spectroscopic galaxy clustering constraints.
}
\label{fig:late_time_comp}
\end{figure*}

\subsection{Comparison With Other Works}

We now compare our constraints on $\Omegam$ and $\sigma_8$ to those from a variety of cosmological probes of late-time matter clustering. These comparisons are summarized in Figure~\ref{fig:late_time_comp}.  The orange contours represent the $68\%$ and $95\%$ confidence levels as obtained without the AbacusSummit prior.

In the top-left panel of Figure \ref{fig:late_time_comp} we compare our results to those from other recent optical cluster studies, namely the DES-Y3 4$\times$2pt+N analysis of \citep[potherurple;][]{DESY3_6x2_et_al_2025}, the KiDS-1000 analysis of AMICO-selected cluster abundances and lensing \citep[brown;][]{Lesci_et_al_2025}, and the analysis of SDSS redMaPPer cluster abundances and clustering with HSC lensing from \citep[light blue;][]{Sunayama_et_al_2024}. We note that 4$\times$2pt+N refers to the combination of large-scale cluster-shear, cluster-galaxy, cluster-cluster, and galaxy-galaxy correlations with cluster abundances (the ``N'').

There is broad agreement between the various optical cluster analyses, though we note the KiDS-1000 contour results in surprisingly low matter densities that are in strong tension with the CMB.  We anticipate that this behavior has a similar origin to that of the original DES-Y1 analysis of \citet{DESY1CL_2020_et_al}.  In particular, while the KiDS-1000 analysis is restricted to relatively small scales ($r_p < 3 \, h^{-1} \, \mathrm{Mpc}$), we find that the impact of cluster selection in \redmapper\ is non-negligible even within the 1-halo term (e.g., right-panel of Fig.~2 in \citet{Salcedo_et_al_2023}).\footnote{While the AMICO cluster finder will result in selection effects that are different in detail from those of redMaPPer, we do anticipate the two will share qualitative similarities.}

Compared to these other optical cluster cosmology constraints, our results are more precise due to a variety of differences in methodology. Specifically, our simulation-based framework enables us to reach lower richness thresholds, which increases constraining power.  Moreover, our pipeline accounts for the cosmology dependence of cluster selection effects while reducing the number of model parameters that significantly impact cluster lensing, leading to improved statistical precision.  Together, these effects enable us to achieve tight cosmological constraints.

In the top right panel of Figure~\ref{fig:late_time_comp} we compare our results to constraints from X-ray and SZ cluster abundance studies: the analysis of SZ-selected clusters in SPT calibrated with DES and HSC lensing \citep[olive;][]{Bocquet_et_al_2024}, the analysis of X-ray clusters in the first eROSITA all sky survey (eRASS1) calibrated with DES, HSC, and KiDS lensing \citep[pink;][]{Ghirardini_et_al_2024}, and the analysis of X-ray clusters in the eROSITA Final Equatorial Depth Survey (eFEDS) calibrated with HSC lensing \citep[dark cyan;][]{Chiu_eFEDS_et_al_2023}. Our constraints from DES-Y1 optical clusters are competitive in precision with those from the X-ray and SZ samples. We find good agreement between our constraints and those from SPT and eFEDS clusters, and a significant $4.1\sigma$ tension with results from X-ray clusters in the eRASS1. The tension between the two eROSITA datasets is $2.0\sigma$. These tensions are driven primarily by the eRASS1 data's preference for a high $\sigma_8$. The exact source of this discrepancy is unclear, though \citet{Ghirardini_et_al_2024} show that their constraints can be impacted by the precise mass function they assume. Their fiducial analysis assumes the mass function parameterization of \citep{Tinker_et_al_2008}, but they show that assuming the parameterizations of either \citep{Tinker_et_al_2010} or \citep{Despali_et_al_2016} shifts their constraints to lower $\sigma_8$, substantially reducing the tension with our work. 

\begin{figure}
\centering \includegraphics[width=0.48\textwidth]{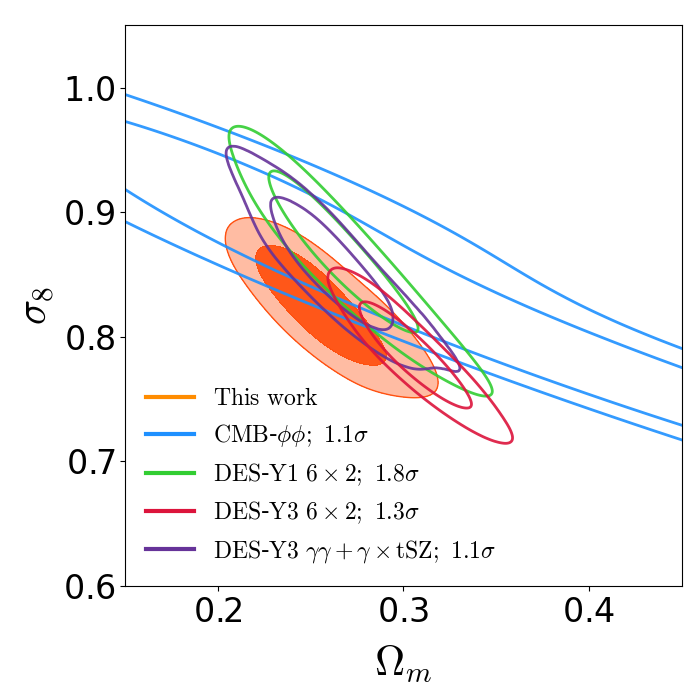}
\caption{68\% and 95\% confidence contours in the $\sigma_8$--$\Omegam$ plane for a variety of different experiments.  Our constraints with and without the AbacusSummit prior are shown in red and orange respectively, and tensions are reported relative to the run without the AbacusSummit prior.  We compare our results to the analysis by \citep[light green;][]{Xu_et_al_2024} which extends the DES-Y1 3$\times$2-point results by including cross-correlations with \textit{Planck} lensing (light green), the analogous analysis in DES-Y3 \citep[red;][]{DESY36x2CMB_et_al_2023}, the DES-Y3 cosmic shear and the shear-tSZ cross-correlation \citep[purple;][]{Pandey_et_al_2025}, and CMB-lensing from the combination of SPT, {\it{Planck}} and ACT data \citep[light blue;][]{Qu_SPA_lens_et_al_2025}.} 
\label{fig:CMB_sec_comp}
\end{figure}

We next turn to comparisons with constraints from the combination of cosmic shear, galaxy-galaxy lensing, and galaxy clustering, typically referred to as $3\times2$.  The bottom left panel in Figure~\ref{fig:late_time_comp} compares our constraints with $3\times2$. from DES-Y3 \citep[green;][]{DES_3x2pt_2021}, KiDS{-}1000 \citep[blue;][]{Heymans_et_al_2021} and HSC-Y3 \citep[gold;][]{ZhangTQ_HSC3x2_et_al_2025}. We can see our constraints are both in agreement with and of comparable precision to those from Stage-III 3$\times$2-pt analyses. 

The bottom right panel of Figure~\ref{fig:late_time_comp} compares our constraints with several full-shape spectroscopic galaxy clustering analyses, namely DESI-DR1 \citep[dark pink;][]{DESIDR1_FS_et_al_2024}, the effective field theory based analysis of the BOSS galaxy power-spectrum and bispectrum of \citep[light green][]{Philcox_et_al_2022}, the redshift-space galaxy clustering analysis from the BACCO team \citep[gray;][]{Pellejero-Ibanez_et_al_2024}, and the DarkEmulator based analysis of the BOSS redshift-space galaxy clustering \citep[orange;][]{Miyatake_et_al_2022}. We find we are in good agreement with these full-shape clustering analyses, with the exception of a $2.4\sigma$ tension with \citep{Philcox_et_al_2022} due to our preference for a lower matter density, and their preference for a lower $\sigma_8$.  

We further compare our results against cosmological constraints derived using CMB secondaries, as illustrated in Figure~\ref{fig:CMB_sec_comp}.  The analyses we consider are the extension of the DES-Y1 3$\times$2-pt analysis to cross-correlations with CMB secondaries \citep[light green;][]{Xu_et_al_2024}, this same extension for DES-Y3 \citep[red;][]{DESY36x2CMB_et_al_2023}, the combination of DES-Y3 cosmic shear and the cross-correlation between shear and thermal SZ observations from ACT found in \citep[purple;][]{Pandey_et_al_2025}, and CMB-lensing posterior obtained by combining SPT, {\it{Planck}} and ACT data \citep[light blue;][]{Qu_SPA_lens_et_al_2025}.  We see our results are in good agreement with all four measurements while also achieving comparable precision.

\section{Conclusions}
\label{sec:conclusions}

We have presented a novel simulation-based forward-modeling framework for analyzing cluster abundance and weak-lensing data.  Our approach explicitly incorporates the cosmology-dependent impact of projection effects in galaxy clustering while still enabling us to marginalize over cluster miscentering and baryonic effects. We have applied this framework to redMaPPer clusters selected in DES-Y1 data and show that our methodology successfully describes the lensing profiles of these clusters down to small scales ($0.2 \, h^{-1} \, \mathrm{Mpc}$) and low richness ($\lambda = 10$), with a goodness-of-fit metric of $\chi^2/\mathrm{dof}=190.51/188$.

From this analysis, we obtain constraints on the cosmological matter content and clustering: $\Omegam = 0.254 \pm 0.023$, $\sigma_8 = 0.826 \pm 0.032$, and $S_8 = 0.758 \pm0.20$. About $10\%$ of the points in our MCMC chain lie in a region of parameter space not covered by the AbacusSummit simulations used to train our emulator, extending to lower values of $\Omegam$. However, imposing a prior that eliminates models outside the AbacusSummit range has only a small impact on our parameter estimates. When combining our cluster constraints with DES-Y1 $3\times2$ constraints we obtain $\Omegam = 0.266\pm0.016$, $\sigma_8 = 0.815\pm0.025$, and $S_8 = 0.768\pm0.15$.  These combined results require negligible model extrapolation, with $\approx 99.3\%$ of the points in the chain contained within the AbacusSummit prior.

Our constraints from DES-Y1 clusters are competitive in precision and in mostly good agreement with literature constraints from other low-redshift growth probes, including Stage III 3$\times$2-pt analyses, full-shape spectroscopic clustering constraints, and CMB-secondaries as summarized in Figures~\ref{fig:late_time_comp} and~\ref{fig:CMB_sec_comp}. Compared to other optical cluster cosmology analyses, our results are more precise.  This is partly due to the inclusion of information from small scales and low richness, and partly because our forward modeling framework characterizes the cosmology dependence of cluster projection effects while reducing the number of free parameters that significantly impact the predicted weak lensing signal.  

Our constraints agree with those from SZ clusters, weak lensing, and full-shape galaxy clustering. They disagree (at $4.1\sigma$) with those from \citep{Ghirardini_et_al_2024} based on X-ray clusters from eRASS1, which imply significantly higher $\sigma_8$ at a given $\Omegam$. In contrast, we find our constraints to be in agreement with eFEDS X-ray clusters, which are themselves in $2.0\sigma$ tension with X-ray clusters in eRASS1. The source of these discrepancies is currently unclear.  Despite the complications of projection effects, we consider our analysis of optically selected clusters to be as robust as those from SZ and X-ray clusters because our simulation tests show that we can model optical selection directly.  Critically, this modeling of selection removes mass-observable scatter as an adjustable parameter degenerate with cosmological parameters (see \citet{Wu_et_al_2021}).

Compared to the original DES-Y1 cluster analysis of \citet{DESY1CL_2020_et_al}, our results are much closer to those of a {\it Planck}-normalized $\Lambda$CDM cosmology, with the difference being significant at $2.58\sigma$ rather than $5.6\sigma$. The significance of the difference between our results and the combined \textit{Planck}, SPT, and ACT primaries is $2.86\sigma$. In the panoply of low redshift matter clustering constraints from weak lensing and redshift-space distortions, our measurement leans towards lower values of $\Omegam$ and $S_8$.

Our results demonstrate that the forward-modeling framework presented in this paper is capable of robustly characterizing optical cluster selection to enable precise and accurate cosmological analyses from optical cluster samples. The derived constraints are competitive and consistent with a wide field of low-redshift probes of large-scale structure.  Moving forward, we aim to improve the fidelity of our forward-modeling framework, both through improved modeling of the galaxy population, and by modifying the cluster finding process to make it easier to emulate.  Application of this type of improved analyses to more recent DES data and upcoming LSST, \textit{Roman}, and \textit{Euclid} will enable us to deliver on the promise of the use of optical galaxy clusters as a tool for precision cosmology in the Stage~IV era.

\section*{Acknowledgements}
We thank Tomomi Sunayama, Enrique Paillas, and I-Non Chiu for valuable discussions about this work. This work was supported by the ``Maximizing Cosmological Science with the Roman High Latitude Imaging Survey'' Roman Project Infrastructure Team (NASA grant 22-ROMAN11-0011). ANS and ER received funding for this work from the Department of Energy (DOE) grant DE-SC0009913. ANS received funding for this work from DOE grants DE-SC0020247 and DE-SC0025993 as well as the David and Lucile Packard Foundation. HW and SC are supported by the DOE Award DE-SC0010129 and the NSF Award AST-2509910. Simulations were analyzed in part on computational resources of the Ohio Supercomputer Center \citep{OhioSupercomputerCenter1987}, with resources supported in part by the Center for Cosmology and AstroParticle Physics at the Ohio State University. We gratefully acknowledge the use of the {\sc{matplotlib}} software package \citep{Hunter_2007}, the {\sc{GetDist}} software package \citep{Lewis_GetDist_2019}, and the GNU Scientific library \citep{GSL_2009}. This research has made use of the SAO/NASA Astrophysics Data System.

\section*{Data Availability}
The data underlying this article will be shared on reasonable request to the corresponding author.

\bibliographystyle{apsrev}
\bibliography{masterbib}

\begin{appendix}

\section{Robustness to Selection Choices}
\label{app:robustness}

\begin{figure}
\centering \includegraphics[width=0.45\textwidth]{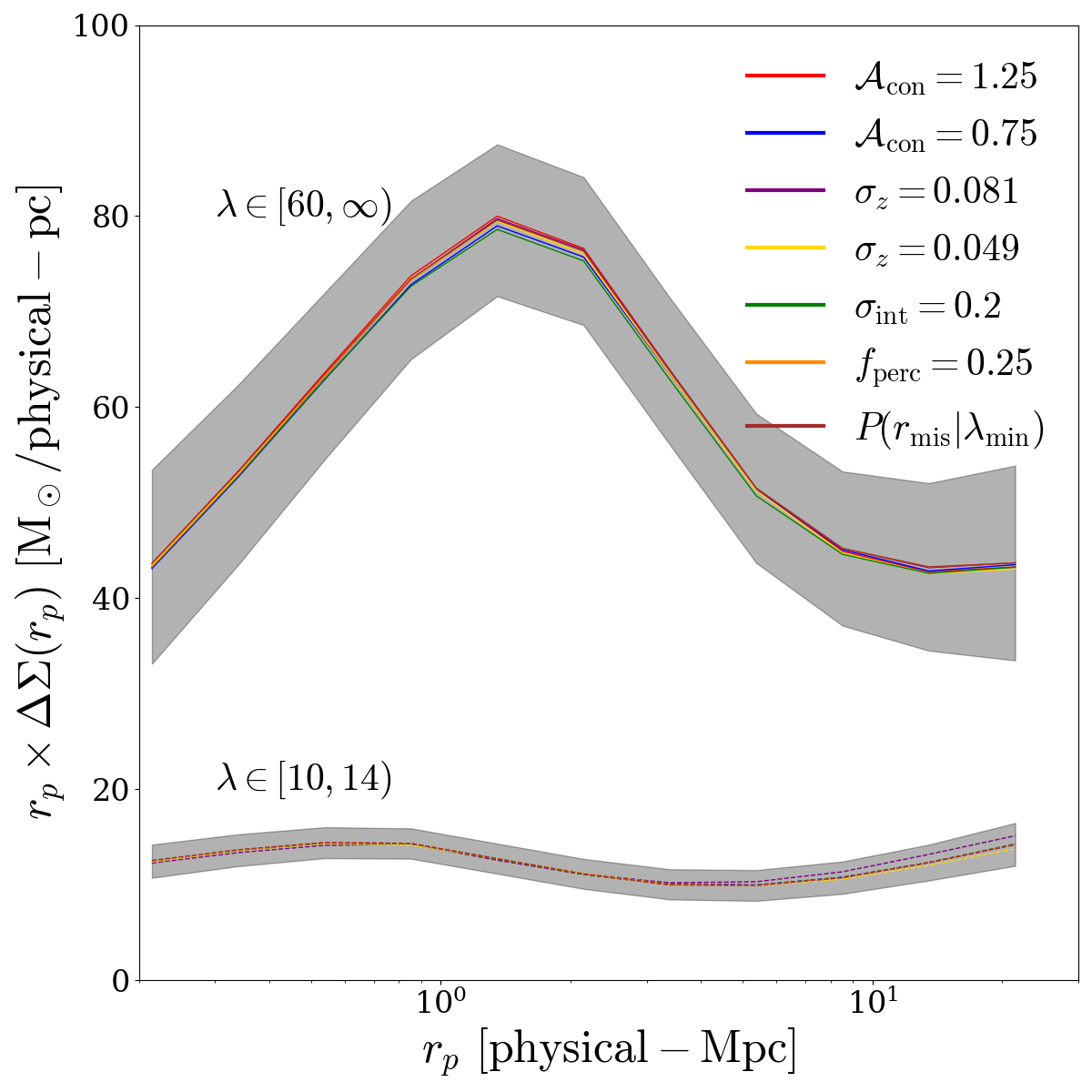}
\caption{We show how our model predictions for low and high richness lensing responds to variations in our fiducial HOD and cylinder selection assumptions. Red and blue lines show variations in the assumed concentration-mass relation, purple and yellow show variations in the assumed depth of the cylinder selection, the green line shows the impact of allowing for super-poissonian scatter in the satellite occupation, the orange line shows the impact of raising our cylinder selection percolation threshold to $25\%$, and the brown line shows the impact of evaluating our miscentering model at each bin's minimum rather than mean richness.} 
\label{fig:robust}
\end{figure}

In Sections~\ref{sec:finding} we identify a variety of simplifying assumptions made in our HOD modeling and mock cluster selection algorithm, specifically: 1) we fix the cylinder depth $\sigma_z$ in each redshift bin to the width of redshift kernel measured using the method of \citet{Costanzi_et_al_2019b}; 2) we assume a fixed concentration-mass relation; 3) we assume that the scatter in the satellite occupation is purely Poisson; 4) we adopt a fixed threshold for percolation in our cluster selection algorithm; and 5) we evaluate the miscentering offset distribution in equation~\ref{eq:mis_dist} at each richness bin's mean richness.

Most of these assumptions are well motivated: 1) in \citet{Salcedo_et_al_2024b} we find $d_\mathrm{cyl}$ to be poorly constrained by cluster lensing, 2) we the impact of variations in the concentration-mass relation to be most important well within the aperture for optical cluster selection, 3) we find in \citet{Salcedo_et_al_2024b} a lack of evidence for super-poissonian scatter from cluster lensing.

In Figure~\ref{fig:robust} we further validate our insensitivity to these choices. We plot our fiducial model with associated $1\sigma$ errors in gray bands for the highest and lowest richness bins at $z\in[0.2,0.35)$.  We compare these model predictions with: 1) variations in the concentration-mass relation where the galaxy concentration
\beq
c^\mathrm{g} = \mathcal{A}_\mathrm{con} \times c^{m},
\eeq
is allowed to vary by $\pm25\%$ (red/blue) relative to that of the matter; 2) $\pm25\%$ variations in the cylinder projection depth $\sigma_z$ (purple/yellow); 3) a case where we include additional lognormal scatter of $\sigma_\mathrm{int} = 0.2$ in the satellite occupation (green); 4) a case in which we raise the percolation threshold to 25\% (orange); and 5) a case where we evaluate our miscentering model in each bin assuming the richness bin's minimum rather than mean richness (brown). In all cases we find our lensing signal prediction to change by much less than the associated statistical uncertainties.

\section{Baryonification}
\label{app:baryon}

\begin{table*}[]
    \centering
    \caption{Summary of parameters of the baryonification model of \citet{Giri_Schneider_2021} that we use to construct our PCA model for the baryonic suppression of the cluster lensing profile as well as hyperparameters of our baryonic feedback model as described in Section~\ref{sec:baryons}. In each case $X_i = m_{x,i} \times B + b_{x,i}.$}
    \begin{tabular}{ccl}
    \hline 
          Parameter &  Training Range/Fixed Value &  Description \\
    \hline 
    $\log M_c$ & $[13.0, 15.0]$ & Characteristic mass scale where gas profile slope shallows. \\
    $\mu$ & $[0.0, 2.0]$ & Mass dependence of gas profile slope. \\
    $\theta_\mathrm{ej}$ & $[2.0,8.0]$ & Gas ejection radius in units of $r_{200b}$\\
    $\gamma$ & $[1.0, 4.0]$ & Normalization of gas profile slope. \\
    $\delta$ & $[3.0, 11.0]$ & Normalization of gas profile slope. \\
    $\eta$ & $[0.325, 0.475]$ & Mass dependence of the stellar fraction in satellite galaxies in ICL. \\
    $\eta_\delta$ & $[0.05, 4.0]$ & Mass dependence of stellar fraction in central galaxy. \\
    \hline
    $m_{a,2}$ & -40.4 & Slope of linear relation between $B$ and $A_2$. \\
    $b_{a,2}$ & -0.41 & Offset of linear relation between $B$ and $A_2$. \\
    $m_{a,1}$ & 543.2 & Slope of linear relation between $B$ and $A_1$. \\
    $b_{a,1}$ & 10.4 & Offset of linear relation between $B$ and $A_1$. \\
    $m_{a,0}$ & -2433.3 & Slope of linear relation between $B$ and $A_0$. \\
    $b_{a,0}$ & -65.1 & Offset of linear relation between $B$ and $A_0$. \\
    $m_{c,2}$ & 0.97 & Slope of linear relation between $B$ and $C_2$. \\
    $b_{c,2}$ & 0.057& Offset of linear relation between $B$ and $C_2$. \\
    $m_{c,1}$ & -25.8 & Slope of linear relation between $B$ and $C_1$. \\
    $b_{c,1}$ &-2.4  & Offset of linear relation between $B$ and $C_1$. \\
    $m_{c,0}$ & 171.3 & Slope of linear relation between $B$ and $C_0$. \\
    $b_{c,0}$ & 21.8 & Offset of linear relation between $B$ and $C_0$. \\
    \hline
    \end{tabular}
    \label{tab:param2}
\end{table*}
    
Following \cite{Giri_Schneider_2021} we model a ``baryonified'' halo profile as,
\beq
\rho_\mathrm{dmb}(r) = \rho_\mathrm{clm}(r) + \rho_\mathrm{gas}(r) + \rho_\mathrm{cga}(r),
\eeq
a sum of collisionless matter (clm), gas (gas), and central galaxy (cga) profiles. These density profiles are normalized such that,
\begin{align}
\int_0^\infty &\left(  \rho_\mathrm{clm}(r) + \rho_\mathrm{gas}(r) + \rho_\mathrm{cga}(r) \right) 4 \pi r^2 dr \nonumber \\
&= \int_0^\infty \rho_{\mathrm{nfw}}(r) 4 \pi r^2 dr = M_\mathrm{tot}
\end{align}
where $\rho_\mathrm{nfw}$ is the truncated NFW \citep{NFW_1997} profile,
\beq
\rho_\mathrm{nfw}(r | M_h, c) = \frac{\rho_0}{\frac{r}{r_s} ( 1 + \frac{r}{r_s})^2} \frac{1}{ (1 + (\frac{r}{r_t})^2)^2}
\eeq
for a halo of mass $M_h$ and concentration $c = r_h / r_s$. The truncation radius is given by $r_t = 4 \times r_h$. 

The central galaxy profile is described by an exponentially truncated power law,
\beq
\rho_\mathrm{cga} (r) = \frac{ M_h f_\mathrm{cga}(M)}{4 \pi^{3/2} R_h r^2} \mathrm{exp}\left[ - \frac{r}{2 R_h} \right],
\eeq
where $f_\mathrm{cga}(M)$ is the fraction of stars in the central galaxy, and $R_h = 0.015 R_\mathrm{vir}$ is the stellar half-light radius. The gas profile is written,
\beq
\rho_\mathrm{gas} \propto \frac{\Omegab / \Omegam - f_\mathrm{star}(M)}{\left[ 1 + 10 \left( \frac{r}{r_\mathrm{vir}}\right) \right]^{\beta(M)} \left[ 1 + \left(\frac{r}{\theta_\mathrm{ej} r_\mathrm{vir}} \right)^\gamma \right]^\frac{\delta - \beta(M)}{\gamma}},
\eeq
where $f_\mathrm{star}(M)$ is the total fraction of stars in halos. The parameter $\beta$ introduces an additional halo-mass dependence,
\beq
\beta(M_c, \mu) = \frac{3(M/M_c)^\mu}{1 + (M / M_c)^\mu}.
\eeq
The initial collisionless matter profile is initially written as.
\beq
\rho_\mathrm{clm}(r) = \left[ \frac{\Omega_\mathrm{dm}}{\Omegam}  + f_\mathrm{sga}(M) \right]\rho_\mathrm{nfw}(r | M, c),
\eeq
where $\rho_\mathrm{nfw}(r | M, c)$ is the truncated NFW profile with concentrations given by the scaling relation of \citet{Correa_2015} and $f_\mathrm{sga}$ is the stellar fraction from satellite galaxies and intracluster light. This profile is then modified according to the ``adiabatic relaxation'' prescription of \citep{Schneider_et_al_2019} which relates the total mass of the initial NFW profile $M_i$ enclosed within radius $r_i$ to that of the final baryonified profile,
\beq
\frac{r_f}{r_i} - 1 = a \left[ \left( \frac{M_i}{M_f}\right)^n - 1\right],
\eeq
where $a = 0.3$ and $n=2$ as in \citep{Schneider_et_al_2019}, and the enclosed mass $M_i$ and $M_f$ are given by,
\begin{align}
M_i &= M_\mathrm{nfw}(r_i), \nonumber \\
M_f &= f_\mathrm{clm} M_\mathrm{nfw}(r_i) + M_\mathrm{cga}(r_f) + M_\mathrm{gas}(r_f),
\end{align}
where $f_\mathrm{clm} = \Omega_{dm}/\Omegam + f_\mathrm{sga}$. These equations are iteratively solved for,
\beq
\zeta(r_i) = r_f / r_i
\eeq
in order to obtain final collisionless matter profile,
\beq
\rho_\mathrm{clm}(r) = \frac{f_\mathrm{clm}}{4\pi r^2} \frac{d}{dr} M_\mathrm{nfw}\left(r / \zeta(r) \right).
\eeq
Finally we relate the various stellar-to-halo fractions via,
\beq
f_\mathrm{sga}(M) = f_\mathrm{star}(M) - f_\mathrm{cga}(M),
\eeq
where
\beq
f_i(M) = 0.055 \left( \frac{M}{M_s} \right)^{-\eta_i}.
\eeq
where $M_s = 2.5 \times 10^{11} h^{-1} M_\odot$ and $\eta_\mathrm{star} = \eta$ and $\eta_\mathrm{cga} = \eta + \eta_\delta$.

\section{Emulator Extrapolation Tests}
\label{app:extrapolation}

\begin{figure*}
\centering \includegraphics[width=1.0\textwidth]{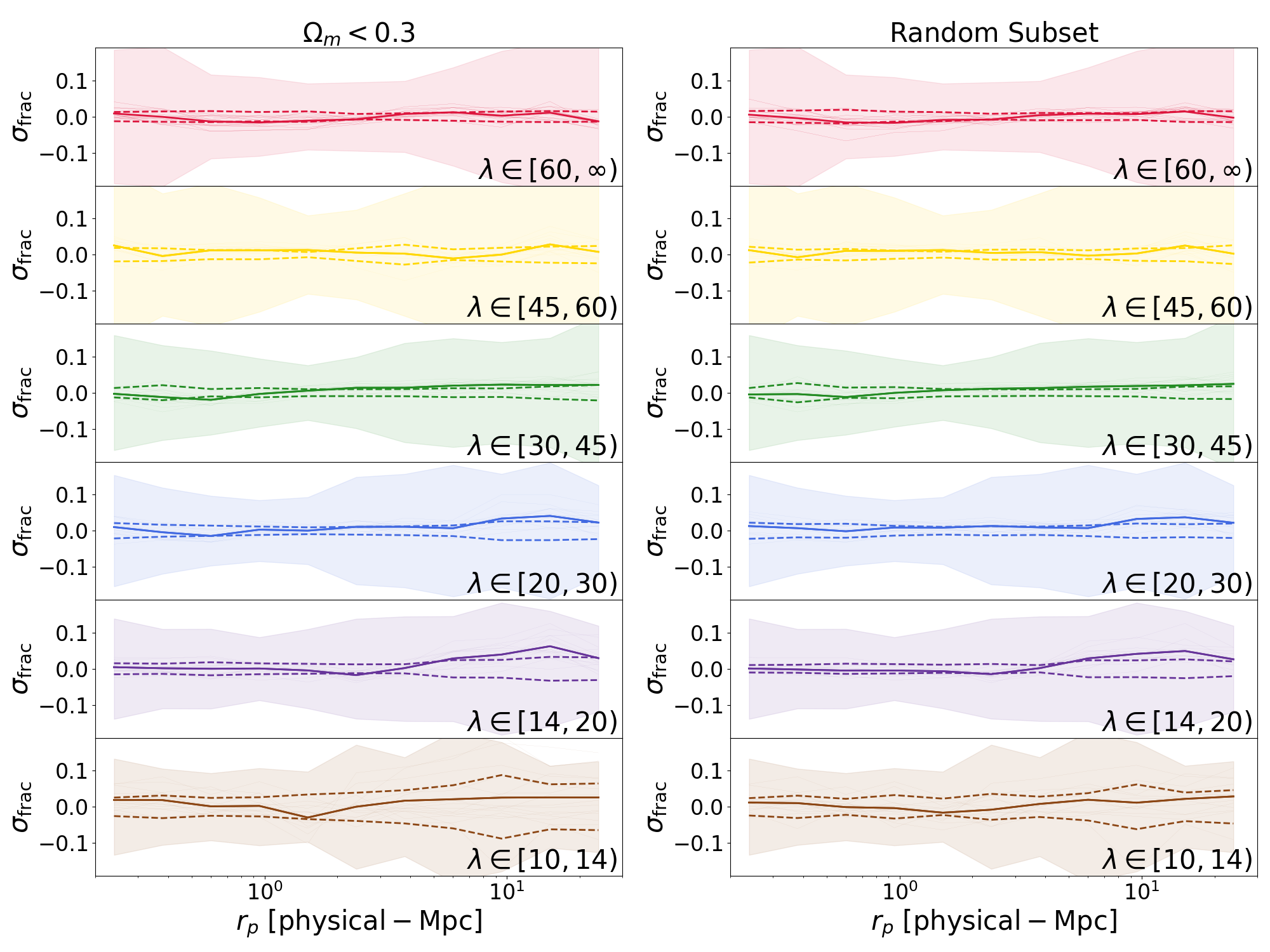}
\caption{Emulator accuracy for an emulator trained on cosmologies with $\Omegam > 0.3$ used to predict lensing profiles of cosmologies $\Omegam < 0.3$ (left) and an emulator removing a equivalent random number of cosmologies used to predict the lensing profiles of those omitted cosmologies (right). In each panel we plot the mean and $1\sigma$ error as solid and dashed lines respectively, and compare to the observable covariance (band).} 
\label{fig:extrapolation}
\end{figure*}

In Section~\ref{sec:results} we present fiducial results with and without the inclusion of the AbacusSummit prior, which is meant to be a conservative ($\approx 6{-}8\sigma$) estimate of the posteriors from CMB+LSS circa 2021. In either case our best-fit parameters lie comfortably within this sampling range. However when we do not impose this prior our sampling extends beyond the AbacusSummit sampling range, and therefore requires some amount of model extrapolation. We find $\sim11\%$ of the points in our chain with no AbacusSummit prior require extrapolation. This extrapolation is not strictly confined to a simple parameter but is most significant in terms of $\Omegam$. Points outside of the sampling range lay within the range of values of $\sigma_8$ supported by AbacusSummit but extend to lower $\Omegam$. 

To test the impact of this extrapolation on our results we define two subsets of our simulation suite, the 15 simulations with $\Omegam < 0.3$ and a separate set of 15 randomly selected simulations. We retrain our emulator with each individual subset of the simulation suite removed. For each of these sets we also compute our lensing datavector at a fiducial HOD model lying at the center of our emulation ranges. We make this choice because we are interested in specifically testing the impact of extrapolation in our cosmology parameters, and because the HOD parameters dominate the inaccuracy in our full emulator.

Figure~\ref{fig:extrapolation} shows the accuracy with which each retrained emulator predicts the lensing signal of the omitted simulations in our lowest redshift bin $z\in[0.20, 0.35)$. In each column, different panels show the fractional accuracy of a different richness bin. The solid lines shows the mean bias, dashed lines show the mean dispersion, and colored bands show the fractional uncertainty from our observable covariance. The left-hand column shows the accuracy with which the emulator trained on all simulations with $\Omegam > 0.3$ can predict the lensing of those simulations with $\Omegam < 0.30$, and the right-hand column shows the accuracy that the emulator trained without a random subset of cosmologies can predict the lensing signal of those cosmologies.  We see that in both cases our emulator prediction is accurate relative to the uncertainty in DES-Y1, and that performance of the two emulators is reasonably consistent. This suggests that limited extrapolation below $\Omegam \approx 0.25$ does not significantly impact our results.  We have also carried out this test with the lensing profiles for the other two redshift bins, finding comparable results (not shown).

\end{appendix}

\end{document}